\documentclass[%
superscriptaddress,
twocolumn,
showpacs,preprintnumbers,
amsmath,amssymb,
aps,
]{revtex4-1}

\usepackage{graphicx}
\usepackage{dcolumn}
\usepackage{bm}
\usepackage{multirow}
\usepackage[mathlines]{lineno}
\usepackage{pifont}

\usepackage{hyperref}
\hypersetup{
    colorlinks=true,
    linkcolor=blue,
    filecolor=blue,
    urlcolor=blue,
    citecolor=blue,
}

\usepackage{color}

\begin{document}

\title{Superexchange interactions and magnetic anisotropy in MnPSe$_3$ monolayer}

\author{Guangyu Wang}
\thanks{These authors contributed equally to this work.}
\affiliation{Laboratory for Computational Physical Sciences (MOE),
	State Key Laboratory of Surface Physics, and Department of Physics,
	Fudan University, Shanghai 200433, China}
\affiliation{Shanghai Qi Zhi Institute, Shanghai 200232, China}

\author{Ke Yang}
\thanks{These authors contributed equally to this work.}
\affiliation{College of Science, University of Shanghai for Science and Technology,
       Shanghai 200093, China}
\affiliation{Laboratory for Computational Physical Sciences (MOE),
	State Key Laboratory of Surface Physics, and Department of Physics,
	Fudan University, Shanghai 200433, China}

\author{Yaozhenghang Ma}
\affiliation{Laboratory for Computational Physical Sciences (MOE),
	State Key Laboratory of Surface Physics, and Department of Physics,
	Fudan University, Shanghai 200433, China}
\affiliation{Shanghai Qi Zhi Institute, Shanghai 200232, China}

\author{Lu Liu}
\affiliation{Laboratory for Computational Physical Sciences (MOE),
	State Key Laboratory of Surface Physics, and Department of Physics,
	Fudan University, Shanghai 200433, China}
\affiliation{Shanghai Qi Zhi Institute, Shanghai 200232, China}

\author{Di Lu}
\affiliation{Laboratory for Computational Physical Sciences (MOE),
	State Key Laboratory of Surface Physics, and Department of Physics,
	Fudan University, Shanghai 200433, China}
\affiliation{Shanghai Qi Zhi Institute, Shanghai 200232, China}

\author{Yuxuan Zhou}
\affiliation{Laboratory for Computational Physical Sciences (MOE),
	State Key Laboratory of Surface Physics, and Department of Physics,
	Fudan University, Shanghai 200433, China}
\affiliation{Shanghai Qi Zhi Institute, Shanghai 200232, China}

\author{Hua Wu}
\thanks{Corresponding author:  wuh@fudan.edu.cn}
\affiliation{Laboratory for Computational Physical Sciences (MOE),
	State Key Laboratory of Surface Physics, and Department of Physics,
	Fudan University, Shanghai 200433, China}
\affiliation{Shanghai Qi Zhi Institute, Shanghai 200232, China}
\affiliation{Collaborative Innovation Center of Advanced Microstructures,
	Nanjing 210093, China}

\date{\today}

\begin{abstract}

Two-dimensional van der Waals magnetic materials are of great current interest for their promising applications in spintronics. In this work, using density functional theory calculations in combination with the maximally localized Wannier functions method and the magnetic anisotropy analyses, we study the electronic and magnetic properties of MnPSe$_3$ monolayer. Our results show that it is a charge transfer antiferromagnetic (AF) insulator. For this Mn$^{2+}$ $3d^5$ system, although it seems straightforward to explain the AF ground state using the direct exchange, we find that the near 90$^\circ$  Mn-Se-Mn charge transfer type superexchange plays a dominant role in stabilizing the AF ground state. Moreover, our results indicate that although the shape anisotropy favors an out-of-plane spin orientation, the spin-orbit coupling (SOC) leads to the experimentally observed in-plane spin orientation. We prove that the actual dominant contribution to the magnetic anisotropy comes from the second-order perturbation of the SOC, by analyzing its distribution over the reciprocal space. Using the AF exchange and anisotropy parameters obtained from our calculations, our Monte Carlo simulations give the N\'eel temperature $T_{\rm N}=47$ K for MnPSe$_3$ monolayer, which agrees with the experimental 40 K. Furthermore, our calculations show that under a uniaxial tensile (compressive) strain, N\'eel vector would be parallel (perpendicular) to the strain direction, which well reproduces the recent experiments. We also predict that $T_{\rm N}$ would be increased by a compressive strain.

\end{abstract}


\maketitle

\section{INTRODUCTION}

Two-dimensional (2D) van der Waals (vdW) materials have attracted widespread attention for several decades due to their highly tunable physical properties and potential applications in multifunctional electronic devices~\cite{42_science2004, 43_Geim2007, 44_RMP2009}.
Recently, with the 2D magnetism observed in atomically thin CrI$_3$ and Cr$_2$Ge$_2$Te$_6$, the 2D vdW magnetic materials have become one of the hot topics in condensed matter research~\cite{Huang_CrI3, Gong_CrGeTe}.
These cleavable materials provide an ideal platform for exploring magnetism in the 2D limit,  where new physics and novel properties are expected.
Moreover, as the vdW interaction between the magnetic layers is usually weak, one may combine these 2D vdW magnetic materials to construct heterostructures, which could produce new properties that are not found in the individual constituents and thus offer immense potential for spintronic applications~\cite{NanolettWang2018, Gibertini2019, LiAM2019, LiangAM2019, ParkPRB2021, Shen_2021_CPL, Gao_2022_CPL}.

\begin{figure}[b]
	\includegraphics[width=8cm]{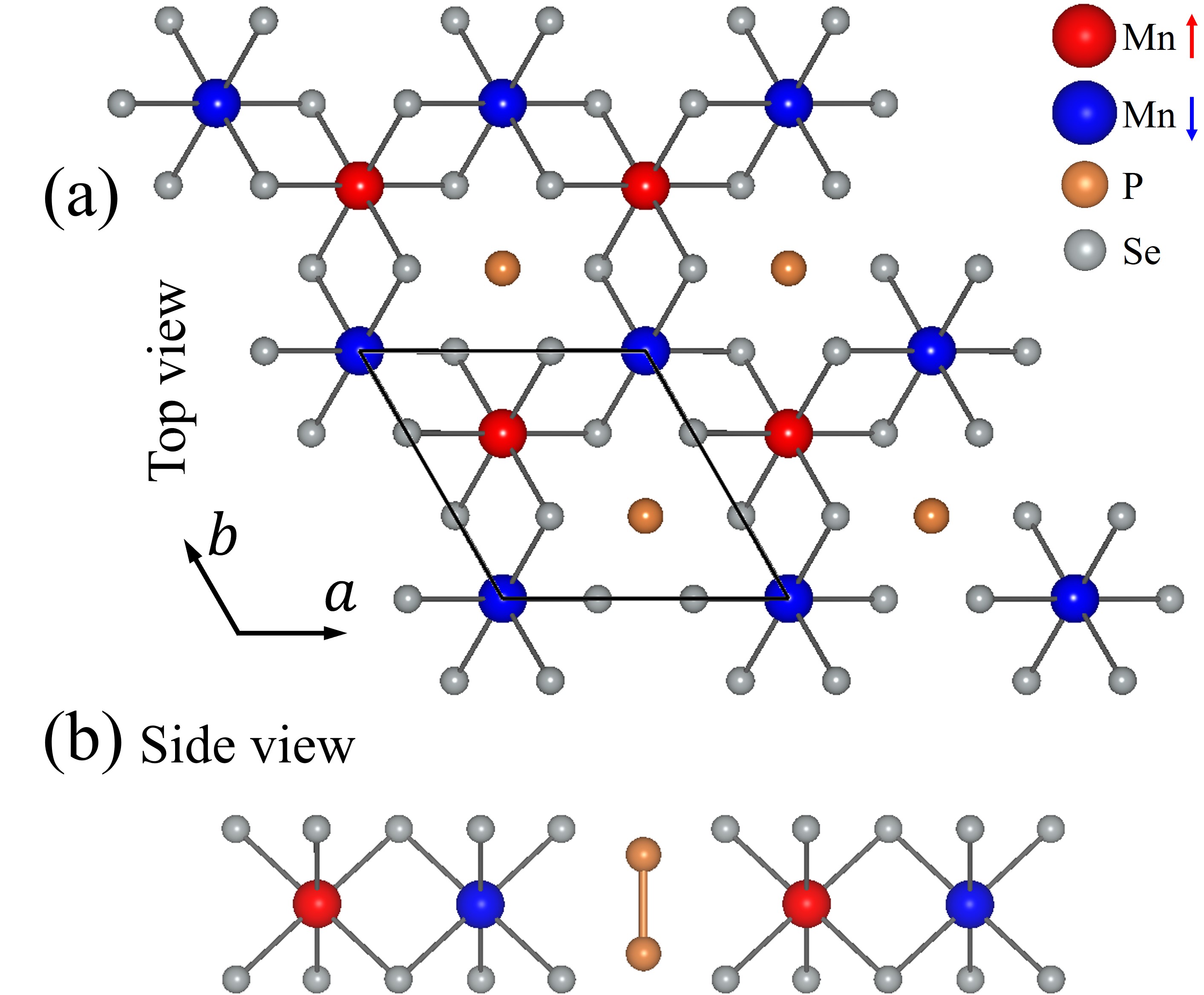}
	\caption {Top view (a) and side view (b) of MnPSe$_3$ monolayer drawn by VESTA~\cite{vesta}, showing the honeycomb Mn ions with up (down) spins, the edge-sharing MnSe$_6$ octahedra, and the P-P dimers.}
	\label{stru}
\end{figure}

One practical strategy to search 2D magnetism is to explore the single layers exfoliated from a bulk vdW material that exhibits robust magnetic order.
The transition metal phosphorous trichalcogenides MPX$_3$ (M = Mn, Ni, Fe, Cd; X= S, Se) are an important 2D vdW magnetic material platform~\cite{39_PRB2015,  41_PRB2016, acsnano2016, Wang2018, 2021_PRB_Liu}.
Among them, MnPSe$_3$ has been extensively studied due to its fascinating and promising properties~\cite{12_1981SSC, 33_XAS, 1_JACS2014, WangJACS2016, SivadasPRL2016, 61_Onga2020, 11_naturenano2021, ref26_PRB, ref37_PRM}.
MnPSe$_3$ crystallizes in the space group R$\overline{3}$ (No. 148), and the honeycomb layers form the rhombohedral ABC stacking along the $c$ axis.
The P-P dimer is located vertically across the center of the honeycomb formed by Mn ions, as seen in Fig. \ref{stru}.
MnPSe$_3$ bulk is a vdW layered AF insulator having the in-plane spins with XY anisotropy and the N\'eel temperature of $T_{\rm N}$=74 K~\cite{12_1981SSC}, and it has the charge-transfer (CT) type optical gap of 2.27 eV~\cite{33_XAS}.
A very recent experimental study showed that MnPSe$_3$ monolayer is dynamically stable and can exist as a 2D freestanding crystal~\cite{11_naturenano2021}, and that it retains the AF ground state with the decreasing $T_{\rm N}$=40 K.
Moreover, it was found that under a uniaxial tensile strain, the N\'eel vector is parallel to the strain direction~\cite{11_naturenano2021}.
So far, however, there has been no systematic theoretical study concerning the magnetic and anisotropic properties of MnPSe$_3$ monolayer, which stimulates our present work.

In this work, we provide insights into the exchange interactions, magnetic anisotropy, and strain effect of MnPSe$_3$ monolayer, using density functional calculations in combination with the maximally localized Wannier functions (MLWFs) method and the magnetic anisotropy analyses. Our calculations show that MnPSe$_3$ monolayer is a CT type insulator with the AF ground state. We demonstrate that the superexchange interaction plays a more important role than the direct exchange in the stabilization of the AF ground state. As for the magnetic anisotropy energy (MAE), which is crucial for 2D magnetism according to the Mermin-Wagner theorem~\cite{Mermin}, our results show that the SOC contributes much more via the second-order perturbation effect to the MAE than the shape anisotropy. We also carefully analyze the distribution of the SOC perturbation energy in the reciprocal space. Using the AF exchange and anisotropy parameters obtained from our calculations, our Monte Carlo simulations show that MnPSe$_3$ monolayer has $T_{\rm N}=47$ K, in agreement with the experimental 40 K~\cite{11_naturenano2021}. Moreover, we find that under a uniaxial tensile (compressive) strain, the N\'eel vector would be parallel (perpendicular) to the strain direction, and the $T_{\rm N}$ would be increased by a compressive strain.

\section{COMPUTATIONAL DETAILS}

Density functional theory (DFT) calculations were carried out using the Vienna $Ab$ $initio$ Simulation Package~\cite{VASP}.
The projector augmented-wave method with a plane-wave basis set was used, and the cut-off energy was set to 400 eV~\cite{PAW}.
The exchange and correlation energy was described by the generalized gradient approximation (GGA) with the Perdew, Burke, and Ernzerhof functional~\cite{PBE}. The lattice parameters of MnPSe$_3$ monolayer are optimized to be $a=b=$ 6.393 \AA, which are almost the same as the experimental bulk value of $a=b=$ 6.387 \AA~\cite{12_1981SSC}.
In the structural optimization, all the atoms were fully relaxed till the total energy converged to 10$^{-6}$ eV and force to 0.01 eV/\AA. To avoid an interaction between the layer and its periodic images, a vacuum of 20 \AA~ was set along the $c$-axis. The Monkhorst-Pack $k$-mesh of 12$\times$12$\times$1 was used for integration over the Brillouin zone~\cite{MP_Method}.
To describe the on-site Coulomb interactions of Mn 3$d$ electrons, the GGA+$U$ method~\cite{UTYPE2} was used with a typical value of the Hubbard $U$ = 4.0 eV~\cite{U_SciAdv_2019, U_npj_2021, U_PRL_2009}.
 As seen below, our calculations with $U$ = 4.0 eV well reproduce the experimental insulating gap, and moreover, the \textcolor{black} {calculated} exchange and anisotropy parameters account for the experimental $T_{\rm N}$ reasonably well.
We also included the SOC effect in our GGA$+U+$SOC calculations to study magnetic anisotropy.
The tight-binding parameters were obtained from MLWFs analysis using the Wannier90 package~\cite{Wannier90, RMP_MLWF}, and they were used to check the relative importance of the direct exchange and the superexchange interactions.
Moreover, we carried out Monte Carlo simulations to estimate the $T_{\rm N}$ of MnPSe$_3$ monolayer, using the Metropolis method~\cite{MC_Metropolis} and a 20$\times$20 spin lattice with a periodic boundary condition.
\textcolor{black} {Such the 20$\times$20 spin lattice turns out to be sufficiently large to ensure a reliable simulation of the magnetic transition temperature.}
During the simulation steps, each spin was rotated randomly in all directions.
At each temperature, we used 3$\times$10$^8$ Monte Carlo steps to reach an equilibrium and 6$\times$10$^8$ steps for a statistical average. The magnetic specific heat at a given temperature was calculated according to
$C_v$=$( \langle E^2 \rangle - \langle E \rangle^2) / (k_{\rm {B}} T^2)$, where $E$ is the total energy of the spin system.

\section{RESULTS AND DISCUSSION}

\begin{figure}[b]
	\includegraphics[width=8.5 cm]{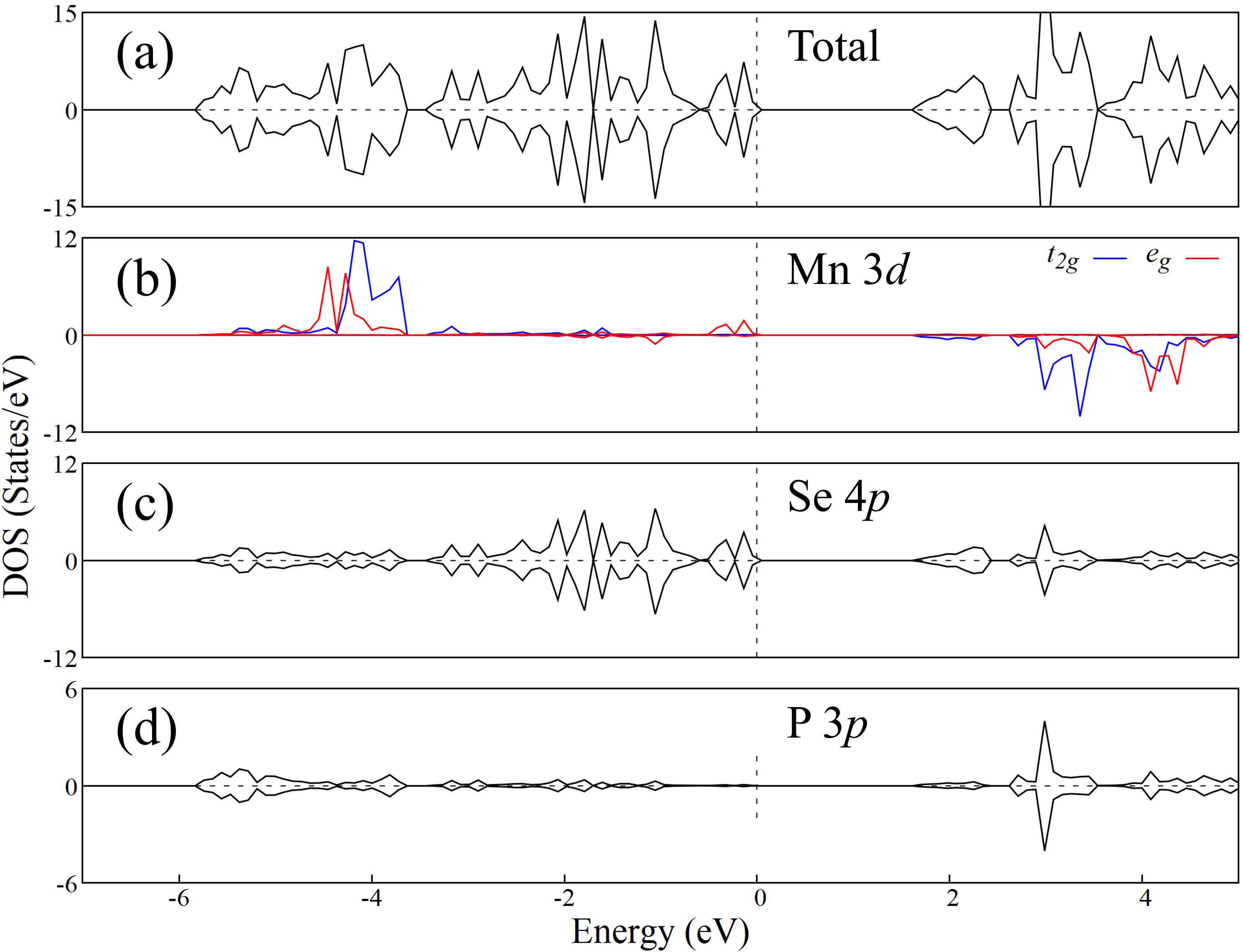}
	\caption {Density of states (DOS) of MnPSe$_3$ monolayer by GGA+$U$. It is an AF insulator with a CT type gap from Se $4p$ to Mn $3d$. The blue (red) curves in (b) stand for the $t_{2g}$ ($e_g$) orbitals. Fermi level is set at zero energy.}
	\label{DOS}
\end{figure}

\subsection{Electronic structure and magnetic exchange interactions}

We performed GGA+$U$ calculations to investigate the electronic structure of MnPSe3 monolayer in the AF insulating ground state. The Mn 3$d$ orbitals in the local MnSe$_6$ octahedral crystal field split into the $t_{2g}$ triplet and $e_g$ doublet.
Our calculations show that the Mn ions each have a local spin moment of 4.6 $\mu_{\rm B}$, indicating the formal Mn$^{2+}$ state with the $t_{2g}^3e_g^2$ ($S$ = 5/2) configuration. The fully occupied up-spin $t_{2g}^3e_g^2$ states are also clearly seen in Fig. \ref{DOS}(b).
A reduction of the spin moment is due to a strong covalency between Mn $3d$ and Se $4p$. Fig. \ref{DOS} shows that the topmost valence bands arise from the Se $4p$ states, and the Mn $3d$ conduction bands lie above the Fermi level by about 2 eV. The strong Mn $3d$-Se $4p$ covalency was also proposed by the recent x-ray photoemission and x-ray absorption spectroscopy~\cite{ref64_PRB2022_masato}, and the charge-transfer (CT) type excitation gap of 2.27 eV from Se $4p$ to Mn $3d$ in the previous optical absorption study~\cite{33_XAS} is here well reproduced by our present calculations.
Therefore, our calculations suggest that MnPSe$_3$ monolayer is a CT insulator.

\begin{figure}[t]
	\includegraphics[width=7cm]{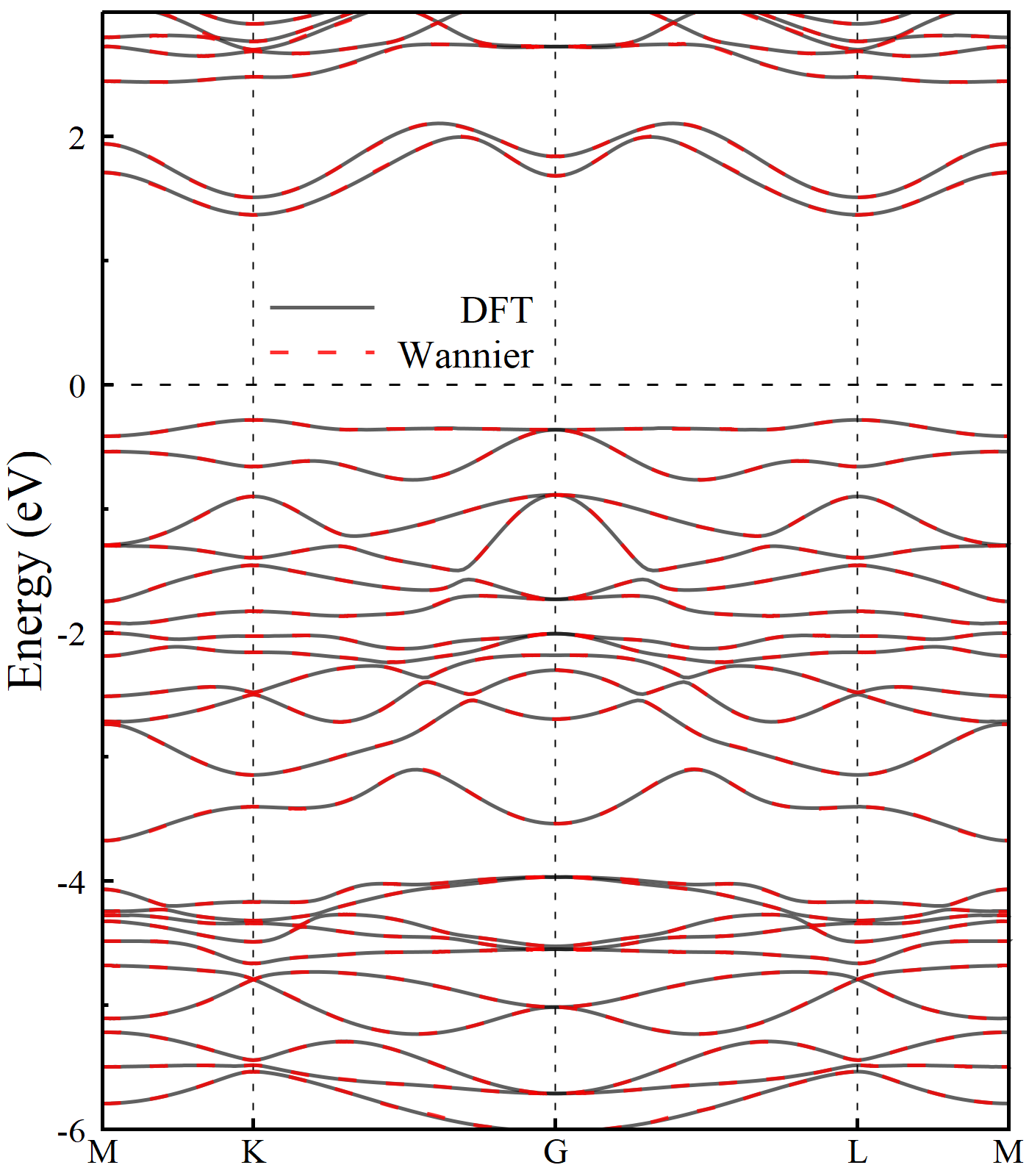}
	\caption {The GGA+$U$ calculated DFT (black solid lines) and Wannier-interpolated (red dashed lines) band structures of MnPSe$_3$ monolayer.}
	\label{bands}
\end{figure}

Now we study the magnetic properties of MnPSe$_3$ monolayer. Our calculations show that the AF N\'eel state is the ground state and it has a lower total energy than the ferromagnetic (FM) state by 26.4 meV per formula unit (fu).
The AF exchange parameter \textcolor{black}{may be} estimated to be 1.41 meV (26.4/3$S^2$) by mapping the energy difference to a Heisenberg spin Hamiltonian \textcolor{black}{($H = \sum_{\emph{i,j}} \frac{\emph{J}}{2} \overrightarrow{S_{i}} \cdot \overrightarrow{S_{j}}$)} for the Mn$^{2+}$ ions ($3d^5$, $S$ = 5/2, counting each pair by $JS^2$) \textcolor{black}{(see more information in the Supplemental Materials (SM))\cite{SM}}. As the Mn$^{2+}$ ions are fully spin polarized, their possible direct exchange can only be AF. Then it may be straightforward to explain the AF ground state of MnPSe$_3$ monolayer simply by the direct exchange. However, the superexchange of neighboring magnetic ions through shared ligands usually plays an important role in determining the magnetic order of transition metal compounds~\cite{Khomskii_2014}.
Therefore, we need to check the relative importance of the direct exchange and superexchange using the MLWFs analyses as detailed below.

\begin{figure}[t]
	\includegraphics[width=9cm]{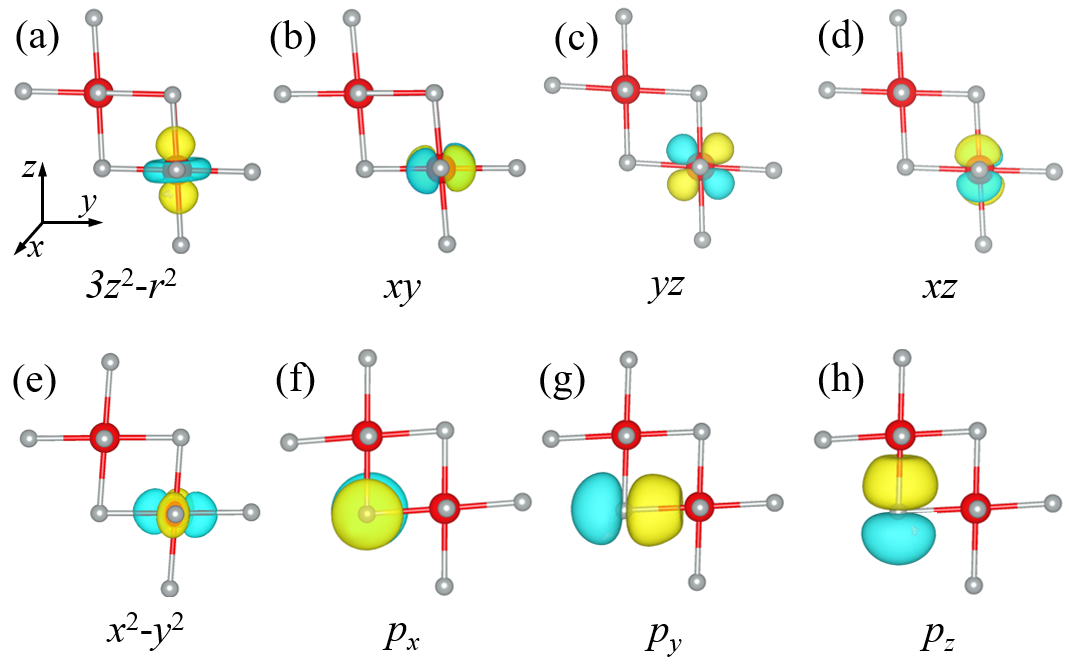}
	\caption {Contour-surface plots of the Mn 3$d$ (a-e) and Se 4$p$ (f-h) Wannier functions. For all plots we choose an isosurface level of $\pm$3.0 (yellow for positive values and cyan for negative values) using the VESTA visualisation program~\cite{vesta}.}
	\label{wannier}
\end{figure}

Direct exchange and superexchange originate from the electronic hopping of Mn-Mn and Mn-Se-Mn, respectively.
To obtain the hopping parameters, we constructed the MLWFs using Mn 3$d$, P 3$s$ and 3$p$, and Se 4$s$ and 4$p$ orbitals as initial guesses.
Fig. \ref{bands} shows the comparison between DFT and Wannier-interpolated band structures of MnPSe$_3$ monolayer, indicating that the chosen MLWFs perfectly reproduce the $ab$ $initio$ electronic states.
The spatial localization is the most important property of the MLWFs, and it is intimately related to the accuracy of the hopping parameters.
To verify the localized features of the MLWFs, we plot the Wannier functions for Mn 3$d$ and Se 4$p$ orbitals in real space.
As shown in Fig. \ref{wannier}, the 3$d$ and 4$p$ Wannier functions are centered on Mn and Se ions, respectively.
Note that we choose the local octahedral $xyz$ coordinate system, with the $xyz$ axes being directed from Mn to neighboring Se ions.
In addition, the shape of these Wannier functions is consistent with the $d$ and $p$ orbitals in isolated atoms.
These results suggest that the Wannier functions are well localized in real space.
Therefore, with the help of these Wannier functions, the hopping integrals of different ions and orbitals can be obtained, and they are crucial for checking the relative importance of the direct exchange and superexchange interactions.

\begin{figure*}[t]
	\includegraphics[height=3.8cm]{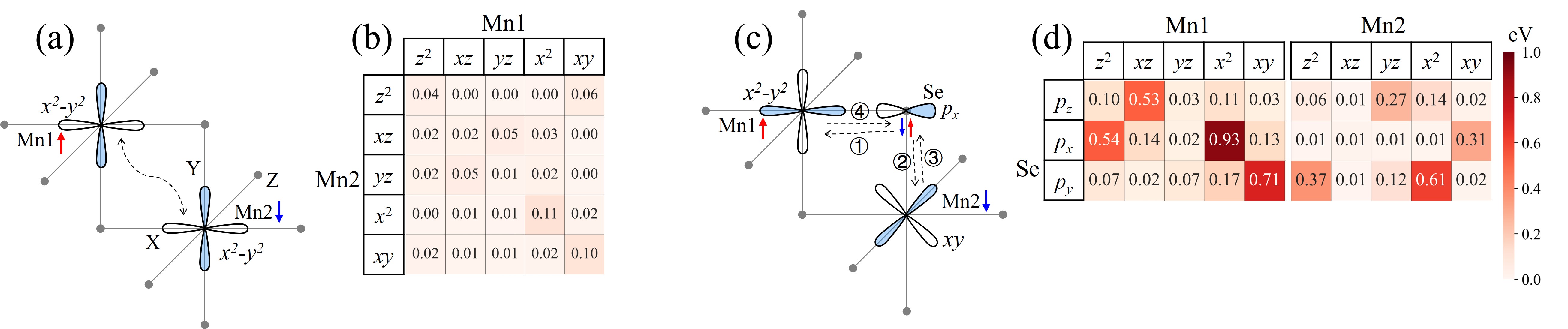}
	\caption {(a): Schematic diagram of the AF direct exchange channel via $d_{x^2-y^2}$ orbitals of Mn1 and Mn2. (b): Hopping integrals of spin-down channels between 3$d$ orbitals of Mn1 and Mn2 calculated by MLWFs basis set. See Fig. S3(a) for the hopping integrals of spin-up channels in the SM~\cite{SM}. (c): The CT type AF superexchange channel via the same $p_x$ orbital. (d): Hopping integrals of spin-down channels between Mn 3$d$ and Se 4$p$ orbitals. See Fig. S3(b) in SM~\cite{SM} for the hopping integrals of spin-up channels. For brevity, we use the notation $d_{x^2}$ for $d_{x^2-y^2}$, and $d_{z^2}$ for $d_{3z^2-r^2}$.}
	\label{hopping}
\end{figure*}

We first investigate the direct exchange in MnPSe$_3$ monolayer.
Fig. \ref{hopping} (a) shows the schematic diagram of the direct exchange.
The direct hopping between two neighboring Mn$^{2+}$ ions is allowed only for spin-antiparallel electrons.
We summarize in Fig. \ref{hopping}(b) the hopping parameters from the  MLWFs calculations, and find that the $d_{x^2-y^2}$ orbitals of two neighboring Mn ions have the largest hopping integral of 0.11 eV and the $d_{xy}$ orbitals the second largest one of 0.10 eV.
All other hopping integrals are about half of them in strength or even less, thus their contribution to the direct exchange is of less concern.

As for the superexchange via Se 4$p$ orbitals in this CT-type insulator, here we focus on the following virtual hopping processes with double holes of Se 4$p$ orbitals in the intermediate excited state after two CT excitations from Se $4p$ to Mn $3d$: $d_i^n p^6 d_j^n \xrightarrow[]{ \bm{1} } d_i^{n+1} p^5 d_j^{n}  \xrightarrow[]{ \bm{2} } d_i^{n+1} p^4 d_j^{n+1}  \xrightarrow[]{ \bm{3}} d_i^{n+1} p^5 d_j^{n}  \xrightarrow[]{ \bm{4} } d_i^n p^6 d_j^n$.
The superexchange channels in MnPSe$_3$ monolayer can be divided into two categories: via different $p$ orbitals or via same $p$ orbitals but with different spins.
As the Se $4p$ electrons on their fat orbitals have negligibly weak Hund exchange, the intermediate double hole states of two different $p$ orbitals either in the spin=1 state or spin=0 one practically have no energy difference, and therefore have no net contribution to the magnetic couplings. In strong contrast, the superexchange channels via same $p$ orbitals allow only AF coupling in MnPSe$_3$. As seen in Fig. \ref{hopping}(c): first, one spin-down $p$-electron from the intermediate Se ion hops to Mn1, then the other spin-up $p$-electron on the same orbital hops to Mn2, and finally these electrons hop back. The former $p_x$-$d_{x^2-y^2}$ hopping is $pd\sigma$ type, and the latter $p_x$-$d_{xy}$ hopping is $pd\pi$ type.
Thus it is not surprising that the former is larger than the latter, see Fig. \ref{hopping}(d).
Moreover, due to the shorter Mn-Se distance of 2.72 \AA (than the Mn-Mn one of 3.70 \AA) and the more expansive Se $4p$ orbital (than the Mn $3d$ one, see Fig. \ref{wannier}), it is evident that the hoppings between Mn and Se ions ($t_{pd}$) in the superexchange are much larger than the above direct $t_{dd}$ hoppings, as seen in Figs. \ref{hopping}(b) and (d). As the strength of the Mn-Mn direct exchange is proportional to $t_{dd}^2$/($U$+4$J_H$), and that of Mn-Se-Mn superexchange is proportional to $t_{pd}^2 t_{pd}^{'2}$/$\Delta ^3$~\cite{Khomskii_2014}, and considering the Hubbard $U$=4 eV and Hund exchange $J_H\approx$ 0.8 eV of the Mn $3d$ electrons and the charge transfer energy $\Delta\approx$ 4.5 eV estimated from our MLWFs calculations, we can infer that the superexchange is much stronger than the direct exchange in MnPSe$_3$ monolayer, for example, the ($x^2-y^2$)-$p_x$-$xy$ superexchange is about 5.0 meV and the ($x^2-y^2$)-($x^2-y^2$) direct exchange is about 1.7 meV.

To summarize, our above results show that MnPSe$_3$ monolayer is a CT type AF insulator, which is in good agreement with the previous experiments~\cite{33_XAS, 11_naturenano2021}.
Although the direct exchange, being absolutely AF for those fully spin-polarized $S$=5/2 Mn$^{2+}$ ions, seems straightforward to explain the AF ground state, we care also the importance of the common superexchange. Then using the MLWFs calculations of the Mn-Mn and Se-Mn hopping parameters, we find that the Mn-Se-Mn superexchange via the strong Mn-Se covalency is more important than the direct Mn-Mn exchange in determining the AF ground state.
\textcolor{black}{Thus, our present results clarify the CT type AF insulating behavior of MnPSe$_3$ monolayer and prove that the superexchange rather than the direct exchange plays a major role in stabilizing the AF ground state.}

\subsection{The MAE and its distributions in reciprocal space}
Now we explore the magnetic anisotropy in MnPSe$_3$ monolayer.
\textcolor{black} {In 2D materials, the magnetic anisotropy breaks the spin rotational invariance and is crucial for stabilizing a long-range magnetic order against a thermal fluctuation.}
The SOC-induced magnetic anisotropy and the magnetic shape anisotropy are two common ones appearing in 2D magnetic materials~\cite{VI3, CrSBr, CrSbSe}.
The SOC is included via the Hamiltonian term $H_{\rm SOC} = \xi \sigma \cdot L$ ($\xi$ is the coupling constant) into our GGA+$U$+SOC calculations.
We assume the spin orientation along the $a$, $b$, and $c$ axes, respectively, and then run their respective self-consistent calculations till the energy difference converges within 2 $\mu$eV/fu.
The results show that MnPSe$_3$ monolayer has the easy $ab$ plane and the $c$ axis spin orientation has a higher energy by 241 $\mu$eV/fu.

\begin{figure}[t]
	\includegraphics[height=2.3cm]{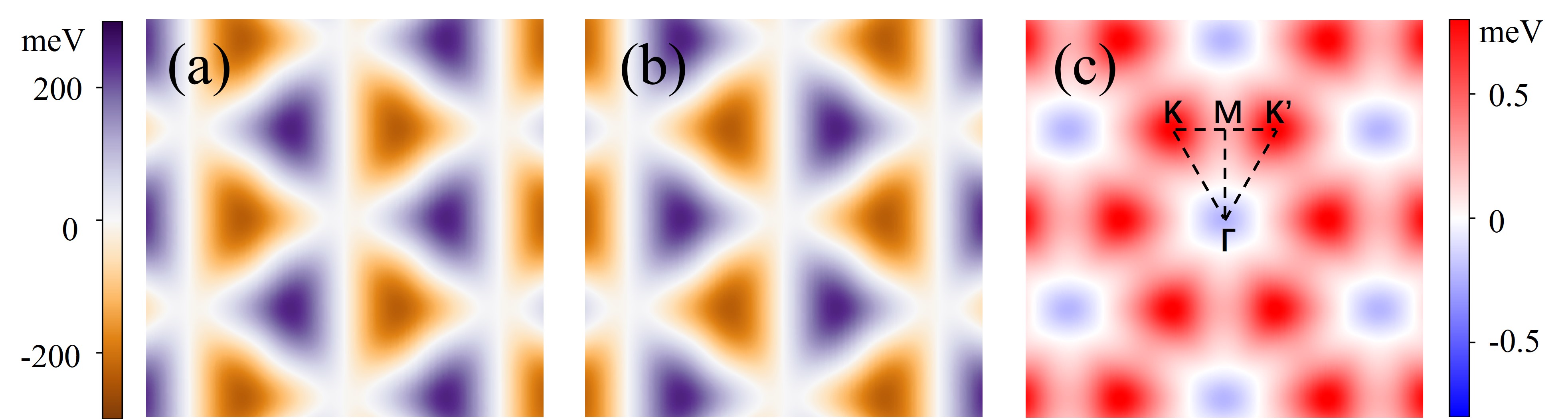}
	\caption {The distributions of MAE in the reciprocal space. (a): $E^{c}-E^{a}$, (b): $E^{-c}-E^{-a}$, (c) $((E^{c}-E^{a})+(E^{-c}-E^{-a}))/2$. See more in the main text.}
	\label{kmae}
\end{figure}

The shape anisotropy energy resulting from the dipole-dipole interaction is
\begin{equation}
E_{\rm dipole-dipole}=\frac{1}{2}\frac{\mu_0}{4\pi}\sum_{i \neq j}^{N} \frac{1}{r^{3}_{ij}}[\vec{M}_{i} \cdot \vec{M}_{j} -
\frac{3}{r^{2}_{ij}}(\vec{M}_{i} \cdot \vec{r}_{ij})(\vec{M}_{j} \cdot \vec{r}_{ij})]
\label{MSA}
\end{equation}
where $\vec{M}_{i}$ represents the Mn$^{2+}$ magnetic moments and $\vec{r}_{ij}$ is a vector connecting the Mn-sites \textit{i} and \textit{j}.
Our results show that the shape anisotropy prefers the easy $c$ axis, and the $ab$ plane spin orientation has a higher energy by 46 $\mu$eV/fu.
Taking into account both the contributions of SOC and shape anisotropy, we find that $c$ axis is the hard magnetization axis with the MAE of 195 $\mu$eV/fu relative to the easy $ab$ plane.
Obviously, the SOC is dominant and it determines the easy plane spin orientation of MnPSe$_3$ monolayer as experimentally measured~\cite{11_naturenano2021}.

As the SOC induced MAE is dominant, here we have a close look at it. The energy gain from SOC according to the perturbation theory is
\begin{equation}
E_{\rm SOC} = \langle o | H_{\rm SOC} | o \rangle + \frac{|\langle o | H_{\rm SOC} | u \rangle |^2}{E_o - E_u}
\label{SOC}
\end{equation}
where $| o \rangle$ and $| u \rangle$ stand for occupied and unoccupied states, $E_o$ and $E_u$ are the corresponding energies. 
In MnPSe$_3$, the Mn$^{2+}$ 3$d^5$ state has only the pure spin S=5/2 and leaves no room for orbital degree of freedom.
This indicates the expectation value of the orbital operator is zero, and the first-order perturbation $\langle o | H_{\rm SOC} | o \rangle$ in Eq. \ref{SOC} has no contribution to the MAE.
Then we care about the second-order perturbation energy
$|\langle o | H_{\rm SOC} | u \rangle |^2 / (E_o - E_u)$.
The MAE caused by the second-order perturbation of SOC is~\cite{PhysRevLett.70.869, Laan_1998_JPCM, Qin_Nanoscale, YueJMMM}:
\begin{equation}
\begin{aligned}
 {\rm MAE} &= E_{\rm SOC}^{\perp} - E_{\rm SOC}^{\parallel}\\
& \approx \frac{1}{2}\sum_{\emph{i,k}} \lbrack \varepsilon_{i}^{\perp}(k) \cdot n_{i}^{\perp}(k) - \varepsilon_{i}^{\parallel}(k) \cdot n_{i}^{\parallel}(k)  \rbrack \cdot \omega(k)
\end{aligned}
\label{MAESOC}
\end{equation}
where $\varepsilon_{i}^{\perp}(k)$ and $\varepsilon_{i}^{\parallel}(k)$ are the $i$th band energy at each $k$-point for MnPSe$_3$ monolayer with the $c$ axis and in-plane spin orientations, respectively.
$n_i^{\perp}(k)$ ($n_i^{\parallel}(k)$) and $\omega(k)$ are the corresponding occupation number and the weight of the $k$ point.

If one plots the distributions of MAE in the reciprocal space, as shown in Fig. \ref{kmae} (a), the MAE contributions at the K and K' points ($\sim$200 meV in strength) would be far from the real value of the MAE (being here only 0.2 meV as seen above), which seems surprising at a first glance. Actually, the `giant' MAE contributions at K and K' are due to the first-order perturbation of SOC:
for systems without an inversion symmetry, the first-order perturbation energy is nonzero at specific $k$-points~\cite{PhysRevLett.70.869, Laan_1998_JPCM, Qin_Nanoscale}. As the inversion symmetry of MnPSe$_3$ monolayer is broken by its AF magnetic ordering, the MAE contributions at K and K' points are dominated by the first-order perturbation energy, which reaches up to $\sim$200 meV as seen in Fig. \ref{kmae}(a).
However, these contributions of first-order perturbation at the specified $k$-points ($k_x$, $k_y$) will be cancelled by those at the ($-k_x$, $-k_y$) points with the same spin orientation, or by those at the same $k$-points but with the opposite spin orientation as seen in Fig. \ref{kmae}(b).
In other words, the first-order perturbation has no net contribution to the MAE upon an integration over the whole Brillouin zone. Moreover, one would clearly see the contributions of second-order perturbation to the MAE in the reciprocal space via the expression $[(E^{c}-E^{a})+(E^{-c}-E^{-a})]/2$ as plotted in Fig. \ref{kmae}(c).  Here the MAE contribution at each $k$-point is indeed tenths of meV in the energy scale, which is in line with the above computed SOC MAE of 241 $\mu$eV/fu.

For a comparison, we also plot the distribution of the MAE of MnPSe$_3$ monolayer in the FM state over the reciprocal space, see Fig. S4 in SM~\cite{SM}.
As the inversion symmetry is preserved in the FM state, the first-order perturbation effect of the SOC is zero at each $k$ points for the MAE~\cite{Laan_1998_JPCM}, and the distributions of MAE in the reciprocal space show only the contributions from the second-order perturbation effects of the SOC, which have the same energy scale of tenths of meV as in the above AF ground state. To summarize, in MnPSe$_3$ monolayer, SOC has a significantly greater contribution to MAE than the shape anisotropy, and the SOC determines the experimental in-plane spin orientation~\cite{11_naturenano2021}. In addition, as the $S$=5/2 Mn$^{2+}$ ion in MnPSe$_3$ monolayer has no orbital moment in its ground state, the MAE arises from the second-order perturbation effects of the SOC. Moreover, in order to clearly see the distribution of the MAE over the reciprocal space, one needs to eliminate the first-order perturbation effect of the SOC which is nonzero at specific $k$ points (but counteracted at $-k$ points) due to the inversion symmetry breaking by the ground state AF order, as shown in Fig. \ref{kmae}.

\subsection{N\'eel temperature and spin orientation under uniaxial strain}

\begin{figure}[t]
	\includegraphics[width=8.5cm]{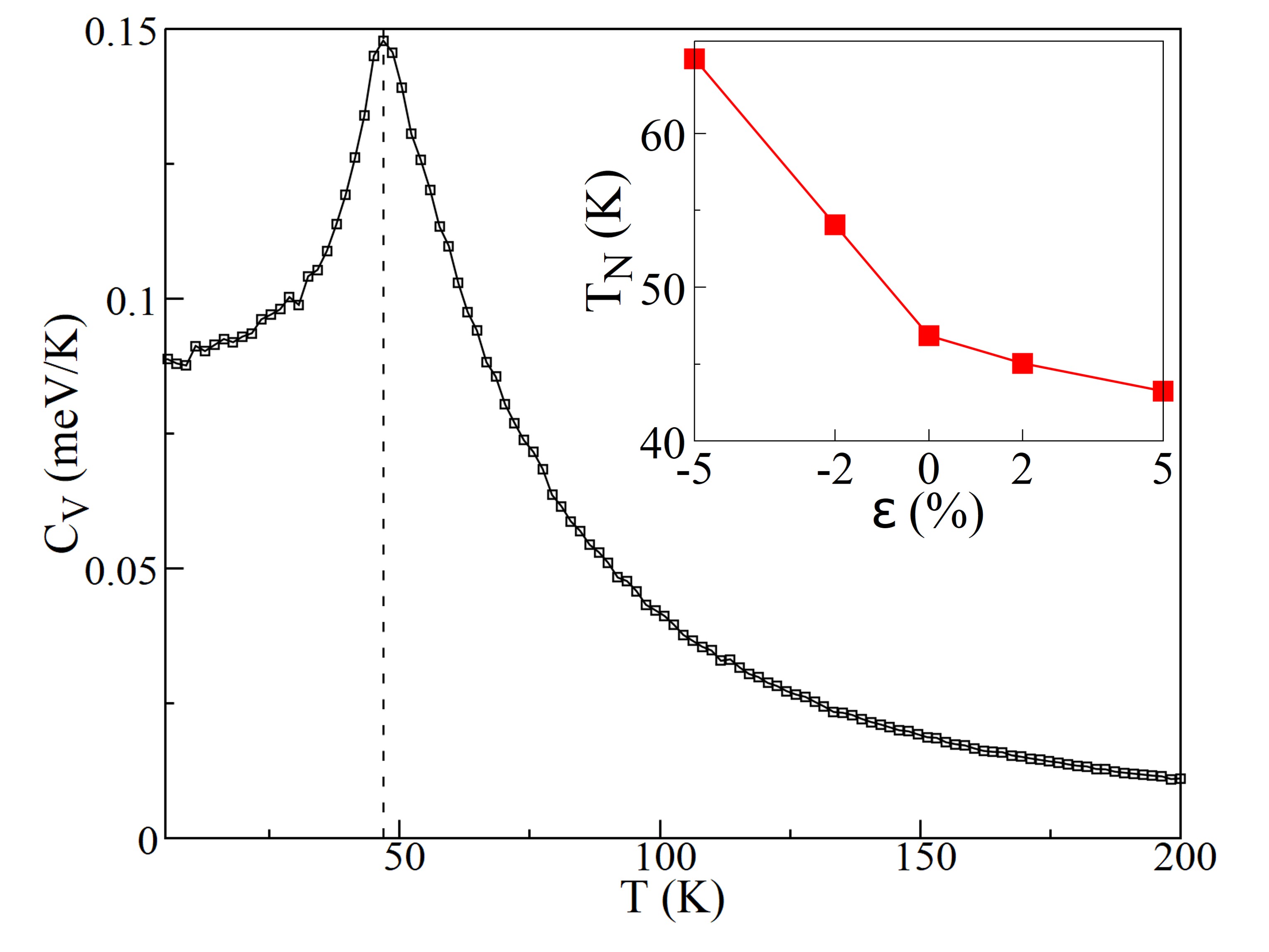}
	\caption {Monte Carlo simulations of the magnetic specific heat of MnPSe$_3$ monolayer. The inset shows the N\'eel temperatures under different uniaxial strains.}
	\label{mc}
\end{figure}

We carry out Monte Carlo simulations to estimate the N\'eel temperature of MnPSe$_3$ monolayer, using the following spin Hamiltonian:
\begin{equation}
\begin{aligned}
H = \sum_{\emph{i,j}} \frac{\emph{J}}{2} \overrightarrow{S_{i}} \cdot \overrightarrow{S_{j}} &+  \sum_{i}\lbrace D(S_{i}^{c})^{2} + E_{n} [(S_{i}^{a})^{2} - (S_{i}^{b})^{2}] \rbrace \\
\end{aligned}
\end{equation}
where the first term stands for the isotropic Heisenberg exchange (AF when $J >$ 0), the second term $D$ describes the longitudinal ($c$ axis) anisotropy, and
the last term $E_n$ represents the transverse ($ab$ plane) anisotropy which would appear  under uniaxial strain (see below).
Using the above GGA+$U$ computed AF exchange parameter $J$=1.41 meV and the anisotropy parameter $D$=31.2 $\mu$eV estimated by the MAE of 195 $\mu$eV per $S$=5/2 Mn$^{2+}$ ion ($D$ = MAE/$S^2$), the calculated magnetic specific heat from our Monte Carlo simulations shows $T_{\rm N}$ = 47 K for MnPSe$_3$ monolayer (see Fig. \ref{mc}), which agrees with the experimental 40 K~\cite{11_naturenano2021}.

\begin{figure}[t]
	\includegraphics[width=8cm]{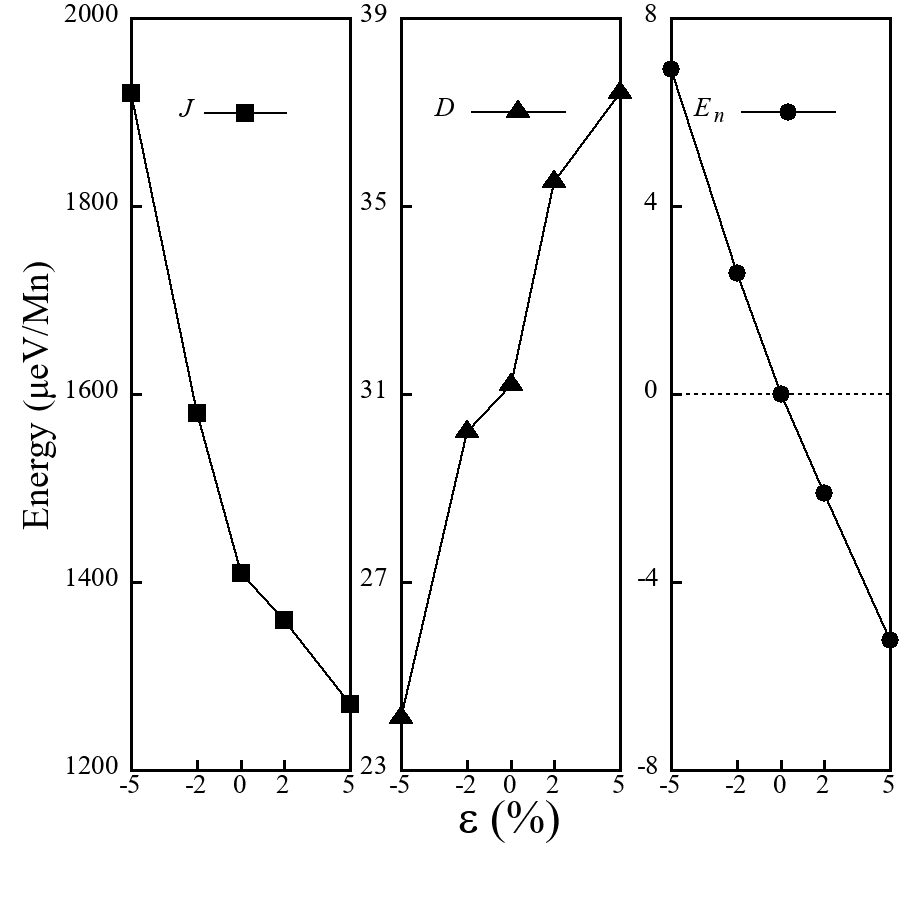}
	\caption {The AF exchange parameter $J$ and the magnetic anisotropy parameters $D$  and $E_n$ for MnPSe$_3$ monolayer under different uniaxial strains along the $a$ axis.}
	\label{JDE}
\end{figure}

\begin{figure}[b]
	\includegraphics[width=7.9cm]{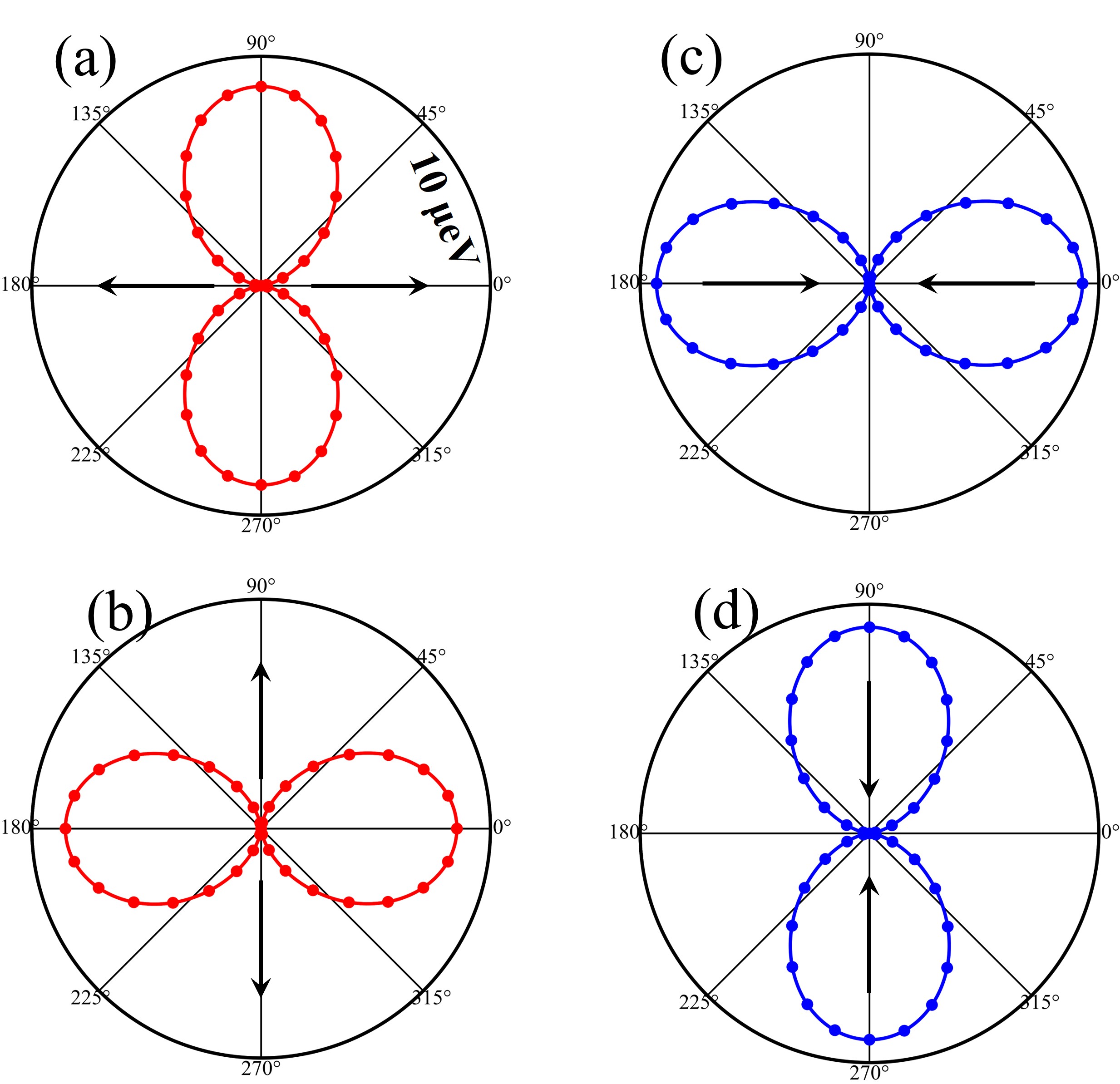}
	\caption {Polar diagrams of the MAE of MnPSe$_3$ monolayer in the $ab$ plane: under 2\% uniaxial tensile strain along 0$^\circ$ (a) and 90$^\circ$ (b) with respect to the $a$ axis, and under 2\% uniaxial compressive strain along 0$^\circ$ (c) and 90$^\circ$ (d) with respect to the $a$ axis.}
	\label{neelvec}
\end{figure}

Strain is an effective tool for tuning various properties of 2D materials.
As seen in Fig. \ref{JDE}, our GGA+$U$+SOC calculations find that the AF exchange parameter $J$ is enhanced under uniaxial compressive strains but is reduced under uniaxial tensile strains, the anisotropy parameter $E_n$ shows a similar tendency, but the anisotropy parameter $D$ shows an opposite tendency. Using those parameters, our Monte Carlo simulations show that the $T_{\rm N}$ of MnPSe$_3$ monolayer increases with the compressive strains but decreases with the tensile strains, as seen in the inset of Fig. \ref{mc}. Moreover, we find that the spin orientation of the Mn$^{2+}$ ions could be tuned by the strains.
As shown in Fig. \ref{neelvec}(a, b), when a 2\% uniaxial tensile strain is applied to MnPSe$_3$ monolayer, the spin orientation (i.e., the N\'eel vector) would be along the direction of the uniaxial strain, which is in good agreement with the experiment~\cite{11_naturenano2021}.
In contrast, the N\'eel vector is perpendicular to the strain direction under 2\% uniaxial compressive strain, as seen in Fig. \ref{neelvec}(c, d).
When the uniaxial tensile (compressive) strain is along 45$^{\circ}$ with respect to the $a$ axis, the N\'eel vector would also be parallel (perpendicular) to the strain direction, see Fig. S5 in SM~\cite{SM}.
The results of 5\% uniaxial strain are similar to the 2\% case except that the in-plane magnetic anisotropy is increased from 10 $\mu$eV to 25 $\mu$eV, see Fig. S6 in SM~\cite{SM}.
To summarize, our GGA+$U$+SOC calculations and Monte Carlo simulations well reproduce the experimental $T_{\rm N}$ of the AF MnPSe$_3$ monolayer and the in-plane spin orientation~\cite{11_naturenano2021}. Moreover, we find that the uniaxial strains can be used to tune the spin orientation, which agrees with the experiment~\cite{11_naturenano2021}, and we predict that the $T_{\rm N}$ would be increased by a compressive strain.

\section{CONCLUSIONS}

To conclude, through density functional theory calculations, magnetic exchange analyses based on MLWFs, and Monte Carlo simulations, we studied the electronic structure and magnetic properties of MnPSe$_3$ monolayer.
Our results show that MnPSe$_3$ monolayer is a CT type AF insulator.
We prove that although the direct exchange between two neighboring $S$=5/2 Mn$^{2+}$ ions is certainly an AF type, the experimental AF order is actually dominated by the near 90$^\circ$ Mn-Se-Mn superexchange via the strong Mn $3d$-Se $4p$ covalency.
We also find that the SOC via the second-order perturbation effects overwhelms the shape anisotropy in yielding the MAE and thus determines the experimental in-plane spin orientation. Moreover, to clearly see the distribution of the MAE over the reciprocal space, one needs to eliminate the first-order perturbation effect of the SOC which is nonzero at specific $k$ points (but counteracted at $-k$ points) due to the inversion symmetry breaking by the AF order.
Using the computed AF exchange parameter and magnetic anisotropy parameters, our Monte Carlo simulations well reproduce the experimental $T_{\rm N}$. Furthermore, we find that a uniaxial strain can be used to tune the spin orientation, which is parallel (perpendicular) to the tensile (compressive) strain direction as the experimentally observed, and we also predict that the $T_{\rm N}$ would be increased by a compressive strain. Thus, this work sheds light on the CT type AF insulating behavior of MnPSe$_3$ monolayer, the major superexchange interactions, magnetic anisotropy and its tuning by strain.

\section{ACKNOWLEDGMENTS}

This work was supported by National Natural Science Foundation of China (Grants No. 12174062, No. 12241402 and No. 12104307).

\bibliography{MnPSe3}

\begin{thebibliography}{49}%
\makeatletter
\providecommand \@ifxundefined [1]{%
 \@ifx{#1\undefined}
}%
\providecommand \@ifnum [1]{%
 \ifnum #1\expandafter \@firstoftwo
 \else \expandafter \@secondoftwo
 \fi
}%
\providecommand \@ifx [1]{%
 \ifx #1\expandafter \@firstoftwo
 \else \expandafter \@secondoftwo
 \fi
}%
\providecommand \natexlab [1]{#1}%
\providecommand \enquote  [1]{``#1''}%
\providecommand \bibnamefont  [1]{#1}%
\providecommand \bibfnamefont [1]{#1}%
\providecommand \citenamefont [1]{#1}%
\providecommand \href@noop [0]{\@secondoftwo}%
\providecommand \href [0]{\begingroup \@sanitize@url \@href}%
\providecommand \@href[1]{\@@startlink{#1}\@@href}%
\providecommand \@@href[1]{\endgroup#1\@@endlink}%
\providecommand \@sanitize@url [0]{\catcode `\\12\catcode `\$12\catcode
  `\&12\catcode `\#12\catcode `\^12\catcode `\_12\catcode `\%12\relax}%
\providecommand \@@startlink[1]{}%
\providecommand \@@endlink[0]{}%
\providecommand \url  [0]{\begingroup\@sanitize@url \@url }%
\providecommand \@url [1]{\endgroup\@href {#1}{\urlprefix }}%
\providecommand \urlprefix  [0]{URL }%
\providecommand \Eprint [0]{\href }%
\providecommand \doibase [0]{http://dx.doi.org/}%
\providecommand \selectlanguage [0]{\@gobble}%
\providecommand \bibinfo  [0]{\@secondoftwo}%
\providecommand \bibfield  [0]{\@secondoftwo}%
\providecommand \translation [1]{[#1]}%
\providecommand \BibitemOpen [0]{}%
\providecommand \bibitemStop [0]{}%
\providecommand \bibitemNoStop [0]{.\EOS\space}%
\providecommand \EOS [0]{\spacefactor3000\relax}%
\providecommand \BibitemShut  [1]{\csname bibitem#1\endcsname}%
\let\auto@bib@innerbib\@empty
\bibitem [{\citenamefont {Novoselov}\ \emph {et~al.}(2004)\citenamefont
  {Novoselov}, \citenamefont {Geim}, \citenamefont {Morozov}, \citenamefont
  {Jiang}, \citenamefont {Zhang}, \citenamefont {Dubonos}, \citenamefont
  {Grigorieva},\ and\ \citenamefont {Firsov}}]{42_science2004}%
  \BibitemOpen
  \bibfield  {author} {\bibinfo {author} {\bibfnamefont {K.~S.}\ \bibnamefont
  {Novoselov}}, \bibinfo {author} {\bibfnamefont {A.~K.}\ \bibnamefont {Geim}},
  \bibinfo {author} {\bibfnamefont {S.~V.}\ \bibnamefont {Morozov}}, \bibinfo
  {author} {\bibfnamefont {D.}~\bibnamefont {Jiang}}, \bibinfo {author}
  {\bibfnamefont {Y.}~\bibnamefont {Zhang}}, \bibinfo {author} {\bibfnamefont
  {S.~V.}\ \bibnamefont {Dubonos}}, \bibinfo {author} {\bibfnamefont {I.~V.}\
  \bibnamefont {Grigorieva}}, \ and\ \bibinfo {author} {\bibfnamefont {A.~A.}\
  \bibnamefont {Firsov}},\ }\href {\doibase 10.1126/science.1102896} {\bibfield
   {journal} {\bibinfo  {journal} {Science}\ }\textbf {\bibinfo {volume}
  {306}},\ \bibinfo {pages} {666} (\bibinfo {year} {2004})}\BibitemShut
  {NoStop}%
\bibitem [{\citenamefont {Geim}\ and\ \citenamefont
  {Novoselov}(2007)}]{43_Geim2007}%
  \BibitemOpen
  \bibfield  {author} {\bibinfo {author} {\bibfnamefont {A.~K.}\ \bibnamefont
  {Geim}}\ and\ \bibinfo {author} {\bibfnamefont {K.~S.}\ \bibnamefont
  {Novoselov}},\ }\href {\doibase 10.1038/nmat1849} {\bibfield  {journal}
  {\bibinfo  {journal} {Nat. Mater.}\ }\textbf {\bibinfo {volume} {6}},\
  \bibinfo {pages} {183} (\bibinfo {year} {2007})}\BibitemShut {NoStop}%
\bibitem [{\citenamefont {Castro~Neto}\ \emph {et~al.}(2009)\citenamefont
  {Castro~Neto}, \citenamefont {Guinea}, \citenamefont {Peres}, \citenamefont
  {Novoselov},\ and\ \citenamefont {Geim}}]{44_RMP2009}%
  \BibitemOpen
  \bibfield  {author} {\bibinfo {author} {\bibfnamefont {A.~H.}\ \bibnamefont
  {Castro~Neto}}, \bibinfo {author} {\bibfnamefont {F.}~\bibnamefont {Guinea}},
  \bibinfo {author} {\bibfnamefont {N.~M.~R.}\ \bibnamefont {Peres}}, \bibinfo
  {author} {\bibfnamefont {K.~S.}\ \bibnamefont {Novoselov}}, \ and\ \bibinfo
  {author} {\bibfnamefont {A.~K.}\ \bibnamefont {Geim}},\ }\href {\doibase
  10.1103/RevModPhys.81.109} {\bibfield  {journal} {\bibinfo  {journal} {Rev.
  Mod. Phys.}\ }\textbf {\bibinfo {volume} {81}},\ \bibinfo {pages} {109}
  (\bibinfo {year} {2009})}\BibitemShut {NoStop}%
\bibitem [{\citenamefont {Huang}\ \emph {et~al.}(2017)\citenamefont {Huang},
  \citenamefont {Clark}, \citenamefont {Navarro-Moratalla}, \citenamefont
  {Klein}, \citenamefont {Cheng}, \citenamefont {Seyler}, \citenamefont
  {Zhong}, \citenamefont {Schmidgall}, \citenamefont {McGuire}, \citenamefont
  {Cobden}, \citenamefont {Yao}, \citenamefont {Xiao}, \citenamefont
  {Jarillo-Herrero},\ and\ \citenamefont {Xu}}]{Huang_CrI3}%
  \BibitemOpen
  \bibfield  {author} {\bibinfo {author} {\bibfnamefont {B.}~\bibnamefont
  {Huang}}, \bibinfo {author} {\bibfnamefont {G.}~\bibnamefont {Clark}},
  \bibinfo {author} {\bibfnamefont {E.}~\bibnamefont {Navarro-Moratalla}},
  \bibinfo {author} {\bibfnamefont {D.~R.}\ \bibnamefont {Klein}}, \bibinfo
  {author} {\bibfnamefont {R.}~\bibnamefont {Cheng}}, \bibinfo {author}
  {\bibfnamefont {K.~L.}\ \bibnamefont {Seyler}}, \bibinfo {author}
  {\bibfnamefont {D.}~\bibnamefont {Zhong}}, \bibinfo {author} {\bibfnamefont
  {E.}~\bibnamefont {Schmidgall}}, \bibinfo {author} {\bibfnamefont {M.~A.}\
  \bibnamefont {McGuire}}, \bibinfo {author} {\bibfnamefont {D.~H.}\
  \bibnamefont {Cobden}}, \bibinfo {author} {\bibfnamefont {W.}~\bibnamefont
  {Yao}}, \bibinfo {author} {\bibfnamefont {D.}~\bibnamefont {Xiao}}, \bibinfo
  {author} {\bibfnamefont {P.}~\bibnamefont {Jarillo-Herrero}}, \ and\ \bibinfo
  {author} {\bibfnamefont {X.}~\bibnamefont {Xu}},\ }\href {\doibase
  10.1038/nature22391} {\bibfield  {journal} {\bibinfo  {journal} {Nature}\
  }\textbf {\bibinfo {volume} {546}},\ \bibinfo {pages} {270} (\bibinfo {year}
  {2017})}\BibitemShut {NoStop}%
\bibitem [{\citenamefont {Gong}\ \emph {et~al.}(2017)\citenamefont {Gong},
  \citenamefont {Li}, \citenamefont {Li}, \citenamefont {Ji}, \citenamefont
  {Stern}, \citenamefont {Xia}, \citenamefont {Cao}, \citenamefont {Bao},
  \citenamefont {Wang}, \citenamefont {Wang}, \citenamefont {Qiu},
  \citenamefont {Cava}, \citenamefont {Louie}, \citenamefont {Xia},\ and\
  \citenamefont {Zhang}}]{Gong_CrGeTe}%
  \BibitemOpen
  \bibfield  {author} {\bibinfo {author} {\bibfnamefont {C.}~\bibnamefont
  {Gong}}, \bibinfo {author} {\bibfnamefont {L.}~\bibnamefont {Li}}, \bibinfo
  {author} {\bibfnamefont {Z.}~\bibnamefont {Li}}, \bibinfo {author}
  {\bibfnamefont {H.}~\bibnamefont {Ji}}, \bibinfo {author} {\bibfnamefont
  {A.}~\bibnamefont {Stern}}, \bibinfo {author} {\bibfnamefont
  {Y.}~\bibnamefont {Xia}}, \bibinfo {author} {\bibfnamefont {T.}~\bibnamefont
  {Cao}}, \bibinfo {author} {\bibfnamefont {W.}~\bibnamefont {Bao}}, \bibinfo
  {author} {\bibfnamefont {C.}~\bibnamefont {Wang}}, \bibinfo {author}
  {\bibfnamefont {Y.}~\bibnamefont {Wang}}, \bibinfo {author} {\bibfnamefont
  {Z.~Q.}\ \bibnamefont {Qiu}}, \bibinfo {author} {\bibfnamefont {R.~J.}\
  \bibnamefont {Cava}}, \bibinfo {author} {\bibfnamefont {S.~G.}\ \bibnamefont
  {Louie}}, \bibinfo {author} {\bibfnamefont {J.}~\bibnamefont {Xia}}, \ and\
  \bibinfo {author} {\bibfnamefont {X.}~\bibnamefont {Zhang}},\ }\href
  {\doibase 10.1038/nature22060} {\bibfield  {journal} {\bibinfo  {journal}
  {Nature}\ }\textbf {\bibinfo {volume} {546}},\ \bibinfo {pages} {265}
  (\bibinfo {year} {2017})}\BibitemShut {NoStop}%
\bibitem [{\citenamefont {Wang}\ \emph
  {et~al.}(2018{\natexlab{a}})\citenamefont {Wang}, \citenamefont {Sapkota},
  \citenamefont {Taniguchi}, \citenamefont {Watanabe}, \citenamefont
  {Mandrus},\ and\ \citenamefont {Morpurgo}}]{NanolettWang2018}%
  \BibitemOpen
  \bibfield  {author} {\bibinfo {author} {\bibfnamefont {Z.}~\bibnamefont
  {Wang}}, \bibinfo {author} {\bibfnamefont {D.}~\bibnamefont {Sapkota}},
  \bibinfo {author} {\bibfnamefont {T.}~\bibnamefont {Taniguchi}}, \bibinfo
  {author} {\bibfnamefont {K.}~\bibnamefont {Watanabe}}, \bibinfo {author}
  {\bibfnamefont {D.}~\bibnamefont {Mandrus}}, \ and\ \bibinfo {author}
  {\bibfnamefont {A.~F.}\ \bibnamefont {Morpurgo}},\ }\href {\doibase
  10.1021/acs.nanolett.8b01278} {\bibfield  {journal} {\bibinfo  {journal}
  {Nano Lett.}\ }\textbf {\bibinfo {volume} {18}},\ \bibinfo {pages} {4303}
  (\bibinfo {year} {2018}{\natexlab{a}})}\BibitemShut {NoStop}%
\bibitem [{\citenamefont {Gibertini}\ \emph {et~al.}(2019)\citenamefont
  {Gibertini}, \citenamefont {Koperski}, \citenamefont {Morpurgo},\ and\
  \citenamefont {Novoselov}}]{Gibertini2019}%
  \BibitemOpen
  \bibfield  {author} {\bibinfo {author} {\bibfnamefont {M.}~\bibnamefont
  {Gibertini}}, \bibinfo {author} {\bibfnamefont {M.}~\bibnamefont {Koperski}},
  \bibinfo {author} {\bibfnamefont {A.~F.}\ \bibnamefont {Morpurgo}}, \ and\
  \bibinfo {author} {\bibfnamefont {K.~S.}\ \bibnamefont {Novoselov}},\ }\href
  {\doibase 10.1038/s41565-019-0438-6} {\bibfield  {journal} {\bibinfo
  {journal} {Nat. Nanotechnol.}\ }\textbf {\bibinfo {volume} {14}},\ \bibinfo
  {pages} {408} (\bibinfo {year} {2019})}\BibitemShut {NoStop}%
\bibitem [{\citenamefont {Li}\ \emph {et~al.}(2019{\natexlab{a}})\citenamefont
  {Li}, \citenamefont {Ruan},\ and\ \citenamefont {Zeng}}]{LiAM2019}%
  \BibitemOpen
  \bibfield  {author} {\bibinfo {author} {\bibfnamefont {H.}~\bibnamefont
  {Li}}, \bibinfo {author} {\bibfnamefont {S.}~\bibnamefont {Ruan}}, \ and\
  \bibinfo {author} {\bibfnamefont {Y.-J.}\ \bibnamefont {Zeng}},\ }\href
  {\doibase https://doi.org/10.1002/adma.201900065} {\bibfield  {journal}
  {\bibinfo  {journal} {Adv. Mater.}\ }\textbf {\bibinfo {volume} {31}},\
  \bibinfo {pages} {1900065} (\bibinfo {year}
  {2019}{\natexlab{a}})}\BibitemShut {NoStop}%
\bibitem [{\citenamefont {Liang}\ \emph {et~al.}(2020)\citenamefont {Liang},
  \citenamefont {Cheng}, \citenamefont {Cui},\ and\ \citenamefont
  {Miao}}]{LiangAM2019}%
  \BibitemOpen
  \bibfield  {author} {\bibinfo {author} {\bibfnamefont {S.-J.}\ \bibnamefont
  {Liang}}, \bibinfo {author} {\bibfnamefont {B.}~\bibnamefont {Cheng}},
  \bibinfo {author} {\bibfnamefont {X.}~\bibnamefont {Cui}}, \ and\ \bibinfo
  {author} {\bibfnamefont {F.}~\bibnamefont {Miao}},\ }\href {\doibase
  https://doi.org/10.1002/adma.201903800} {\bibfield  {journal} {\bibinfo
  {journal} {Adv. Mater.}\ }\textbf {\bibinfo {volume} {32}},\ \bibinfo {pages}
  {1903800} (\bibinfo {year} {2020})}\BibitemShut {NoStop}%
\bibitem [{\citenamefont {Park}\ \emph {et~al.}(2021)\citenamefont {Park},
  \citenamefont {Peng}, \citenamefont {Liang}, \citenamefont {Hallal},
  \citenamefont {Yasin}, \citenamefont {Zhang}, \citenamefont {Song},
  \citenamefont {Kim}, \citenamefont {Kim}, \citenamefont {Weigand},
  \citenamefont {Sch\"utz}, \citenamefont {Finizio}, \citenamefont {Raabe},
  \citenamefont {Garcia}, \citenamefont {Xia}, \citenamefont {Zhou},
  \citenamefont {Ezawa}, \citenamefont {Liu}, \citenamefont {Chang},
  \citenamefont {Koo}, \citenamefont {Kim}, \citenamefont {Chshiev},
  \citenamefont {Fert}, \citenamefont {Yang}, \citenamefont {Yu},\ and\
  \citenamefont {Woo}}]{ParkPRB2021}%
  \BibitemOpen
  \bibfield  {author} {\bibinfo {author} {\bibfnamefont {T.-E.}\ \bibnamefont
  {Park}}, \bibinfo {author} {\bibfnamefont {L.}~\bibnamefont {Peng}}, \bibinfo
  {author} {\bibfnamefont {J.}~\bibnamefont {Liang}}, \bibinfo {author}
  {\bibfnamefont {A.}~\bibnamefont {Hallal}}, \bibinfo {author} {\bibfnamefont
  {F.~S.}\ \bibnamefont {Yasin}}, \bibinfo {author} {\bibfnamefont
  {X.}~\bibnamefont {Zhang}}, \bibinfo {author} {\bibfnamefont {K.~M.}\
  \bibnamefont {Song}}, \bibinfo {author} {\bibfnamefont {S.~J.}\ \bibnamefont
  {Kim}}, \bibinfo {author} {\bibfnamefont {K.}~\bibnamefont {Kim}}, \bibinfo
  {author} {\bibfnamefont {M.}~\bibnamefont {Weigand}}, \bibinfo {author}
  {\bibfnamefont {G.}~\bibnamefont {Sch\"utz}}, \bibinfo {author}
  {\bibfnamefont {S.}~\bibnamefont {Finizio}}, \bibinfo {author} {\bibfnamefont
  {J.}~\bibnamefont {Raabe}}, \bibinfo {author} {\bibfnamefont
  {K.}~\bibnamefont {Garcia}}, \bibinfo {author} {\bibfnamefont
  {J.}~\bibnamefont {Xia}}, \bibinfo {author} {\bibfnamefont {Y.}~\bibnamefont
  {Zhou}}, \bibinfo {author} {\bibfnamefont {M.}~\bibnamefont {Ezawa}},
  \bibinfo {author} {\bibfnamefont {X.}~\bibnamefont {Liu}}, \bibinfo {author}
  {\bibfnamefont {J.}~\bibnamefont {Chang}}, \bibinfo {author} {\bibfnamefont
  {H.~C.}\ \bibnamefont {Koo}}, \bibinfo {author} {\bibfnamefont {Y.~D.}\
  \bibnamefont {Kim}}, \bibinfo {author} {\bibfnamefont {M.}~\bibnamefont
  {Chshiev}}, \bibinfo {author} {\bibfnamefont {A.}~\bibnamefont {Fert}},
  \bibinfo {author} {\bibfnamefont {H.}~\bibnamefont {Yang}}, \bibinfo {author}
  {\bibfnamefont {X.}~\bibnamefont {Yu}}, \ and\ \bibinfo {author}
  {\bibfnamefont {S.}~\bibnamefont {Woo}},\ }\href {\doibase
  10.1103/PhysRevB.103.104410} {\bibfield  {journal} {\bibinfo  {journal}
  {Phys. Rev. B}\ }\textbf {\bibinfo {volume} {103}},\ \bibinfo {pages}
  {104410} (\bibinfo {year} {2021})}\BibitemShut {NoStop}%
\bibitem [{\citenamefont {Shen}\ \emph {et~al.}(2021)\citenamefont {Shen},
  \citenamefont {Tong}, \citenamefont {Hu}, \citenamefont {Zheng},\ and\
  \citenamefont {Duan}}]{Shen_2021_CPL}%
  \BibitemOpen
  \bibfield  {author} {\bibinfo {author} {\bibfnamefont {Y.-H.}\ \bibnamefont
  {Shen}}, \bibinfo {author} {\bibfnamefont {W.-Y.}\ \bibnamefont {Tong}},
  \bibinfo {author} {\bibfnamefont {H.}~\bibnamefont {Hu}}, \bibinfo {author}
  {\bibfnamefont {J.-D.}\ \bibnamefont {Zheng}}, \ and\ \bibinfo {author}
  {\bibfnamefont {C.-G.}\ \bibnamefont {Duan}},\ }\href {\doibase
  10.1088/0256-307X/38/3/037501} {\bibfield  {journal} {\bibinfo  {journal}
  {Chinese Phys. Lett.}\ }\textbf {\bibinfo {volume} {38}},\ \bibinfo {pages}
  {037501} (\bibinfo {year} {2021})}\BibitemShut {NoStop}%
\bibitem [{\citenamefont {Gao}\ \emph {et~al.}(2022)\citenamefont {Gao},
  \citenamefont {Liu}, \citenamefont {Fang}, \citenamefont {Yao}, \citenamefont
  {Wu}, \citenamefont {Xiao}, \citenamefont {Hu}, \citenamefont {Liu},
  \citenamefont {Gao}, \citenamefont {Cao}, \citenamefont {Song}, \citenamefont
  {Meng}, \citenamefont {Chen},\ and\ \citenamefont {Ren}}]{Gao_2022_CPL}%
  \BibitemOpen
  \bibfield  {author} {\bibinfo {author} {\bibfnamefont {R.}~\bibnamefont
  {Gao}}, \bibinfo {author} {\bibfnamefont {C.}~\bibnamefont {Liu}}, \bibinfo
  {author} {\bibfnamefont {L.}~\bibnamefont {Fang}}, \bibinfo {author}
  {\bibfnamefont {B.}~\bibnamefont {Yao}}, \bibinfo {author} {\bibfnamefont
  {W.}~\bibnamefont {Wu}}, \bibinfo {author} {\bibfnamefont {Q.}~\bibnamefont
  {Xiao}}, \bibinfo {author} {\bibfnamefont {S.}~\bibnamefont {Hu}}, \bibinfo
  {author} {\bibfnamefont {Y.}~\bibnamefont {Liu}}, \bibinfo {author}
  {\bibfnamefont {H.}~\bibnamefont {Gao}}, \bibinfo {author} {\bibfnamefont
  {S.}~\bibnamefont {Cao}}, \bibinfo {author} {\bibfnamefont {G.}~\bibnamefont
  {Song}}, \bibinfo {author} {\bibfnamefont {X.}~\bibnamefont {Meng}}, \bibinfo
  {author} {\bibfnamefont {X.}~\bibnamefont {Chen}}, \ and\ \bibinfo {author}
  {\bibfnamefont {W.}~\bibnamefont {Ren}},\ }\href {\doibase
  10.1088/0256-307X/39/12/127301} {\bibfield  {journal} {\bibinfo  {journal}
  {Chinese Phys. Lett.}\ }\textbf {\bibinfo {volume} {39}},\ \bibinfo {pages}
  {127301} (\bibinfo {year} {2022})}\BibitemShut {NoStop}%
\bibitem [{\citenamefont {Momma}\ and\ \citenamefont {Izumi}(2011)}]{vesta}%
  \BibitemOpen
  \bibfield  {author} {\bibinfo {author} {\bibfnamefont {K.}~\bibnamefont
  {Momma}}\ and\ \bibinfo {author} {\bibfnamefont {F.}~\bibnamefont {Izumi}},\
  }\href {\doibase 10.1107/S0021889811038970} {\bibfield  {journal} {\bibinfo
  {journal} {J. Appl. Crystallogr.}\ }\textbf {\bibinfo {volume} {44}},\
  \bibinfo {pages} {1272} (\bibinfo {year} {2011})}\BibitemShut {NoStop}%
\bibitem [{\citenamefont {Wildes}\ \emph {et~al.}(2015)\citenamefont {Wildes},
  \citenamefont {Simonet}, \citenamefont {Ressouche}, \citenamefont {McIntyre},
  \citenamefont {Avdeev}, \citenamefont {Suard}, \citenamefont {Kimber},
  \citenamefont {Lan\ifmmode~\mbox{\c{c}}\else \c{c}\fi{}on}, \citenamefont
  {Pepe}, \citenamefont {Moubaraki},\ and\ \citenamefont {Hicks}}]{39_PRB2015}%
  \BibitemOpen
  \bibfield  {author} {\bibinfo {author} {\bibfnamefont {A.~R.}\ \bibnamefont
  {Wildes}}, \bibinfo {author} {\bibfnamefont {V.}~\bibnamefont {Simonet}},
  \bibinfo {author} {\bibfnamefont {E.}~\bibnamefont {Ressouche}}, \bibinfo
  {author} {\bibfnamefont {G.~J.}\ \bibnamefont {McIntyre}}, \bibinfo {author}
  {\bibfnamefont {M.}~\bibnamefont {Avdeev}}, \bibinfo {author} {\bibfnamefont
  {E.}~\bibnamefont {Suard}}, \bibinfo {author} {\bibfnamefont {S.~A.~J.}\
  \bibnamefont {Kimber}}, \bibinfo {author} {\bibfnamefont {D.}~\bibnamefont
  {Lan\ifmmode~\mbox{\c{c}}\else \c{c}\fi{}on}}, \bibinfo {author}
  {\bibfnamefont {G.}~\bibnamefont {Pepe}}, \bibinfo {author} {\bibfnamefont
  {B.}~\bibnamefont {Moubaraki}}, \ and\ \bibinfo {author} {\bibfnamefont
  {T.~J.}\ \bibnamefont {Hicks}},\ }\href {\doibase 10.1103/PhysRevB.92.224408}
  {\bibfield  {journal} {\bibinfo  {journal} {Phys. Rev. B}\ }\textbf {\bibinfo
  {volume} {92}},\ \bibinfo {pages} {224408} (\bibinfo {year}
  {2015})}\BibitemShut {NoStop}%
\bibitem [{\citenamefont {Lan\ifmmode~\mbox{\c{c}}\else \c{c}\fi{}on}\ \emph
  {et~al.}(2016)\citenamefont {Lan\ifmmode~\mbox{\c{c}}\else \c{c}\fi{}on},
  \citenamefont {Walker}, \citenamefont {Ressouche}, \citenamefont {Ouladdiaf},
  \citenamefont {Rule}, \citenamefont {McIntyre}, \citenamefont {Hicks},
  \citenamefont {R\o{}nnow},\ and\ \citenamefont {Wildes}}]{41_PRB2016}%
  \BibitemOpen
  \bibfield  {author} {\bibinfo {author} {\bibfnamefont {D.}~\bibnamefont
  {Lan\ifmmode~\mbox{\c{c}}\else \c{c}\fi{}on}}, \bibinfo {author}
  {\bibfnamefont {H.~C.}\ \bibnamefont {Walker}}, \bibinfo {author}
  {\bibfnamefont {E.}~\bibnamefont {Ressouche}}, \bibinfo {author}
  {\bibfnamefont {B.}~\bibnamefont {Ouladdiaf}}, \bibinfo {author}
  {\bibfnamefont {K.~C.}\ \bibnamefont {Rule}}, \bibinfo {author}
  {\bibfnamefont {G.~J.}\ \bibnamefont {McIntyre}}, \bibinfo {author}
  {\bibfnamefont {T.~J.}\ \bibnamefont {Hicks}}, \bibinfo {author}
  {\bibfnamefont {H.~M.}\ \bibnamefont {R\o{}nnow}}, \ and\ \bibinfo {author}
  {\bibfnamefont {A.~R.}\ \bibnamefont {Wildes}},\ }\href {\doibase
  10.1103/PhysRevB.94.214407} {\bibfield  {journal} {\bibinfo  {journal} {Phys.
  Rev. B}\ }\textbf {\bibinfo {volume} {94}},\ \bibinfo {pages} {214407}
  (\bibinfo {year} {2016})}\BibitemShut {NoStop}%
\bibitem [{\citenamefont {Du}\ \emph {et~al.}(2016)\citenamefont {Du},
  \citenamefont {Wang}, \citenamefont {Liu}, \citenamefont {Hu}, \citenamefont
  {Utama}, \citenamefont {Gan}, \citenamefont {Xiong},\ and\ \citenamefont
  {Kloc}}]{acsnano2016}%
  \BibitemOpen
  \bibfield  {author} {\bibinfo {author} {\bibfnamefont {K.-z.}\ \bibnamefont
  {Du}}, \bibinfo {author} {\bibfnamefont {X.-z.}\ \bibnamefont {Wang}},
  \bibinfo {author} {\bibfnamefont {Y.}~\bibnamefont {Liu}}, \bibinfo {author}
  {\bibfnamefont {P.}~\bibnamefont {Hu}}, \bibinfo {author} {\bibfnamefont
  {M.~I.~B.}\ \bibnamefont {Utama}}, \bibinfo {author} {\bibfnamefont {C.~K.}\
  \bibnamefont {Gan}}, \bibinfo {author} {\bibfnamefont {Q.}~\bibnamefont
  {Xiong}}, \ and\ \bibinfo {author} {\bibfnamefont {C.}~\bibnamefont {Kloc}},\
  }\href {\doibase 10.1021/acsnano.5b05927} {\bibfield  {journal} {\bibinfo
  {journal} {ACS Nano}\ }\textbf {\bibinfo {volume} {10}},\ \bibinfo {pages}
  {1738} (\bibinfo {year} {2016})}\BibitemShut {NoStop}%
\bibitem [{\citenamefont {Wang}\ \emph
  {et~al.}(2018{\natexlab{b}})\citenamefont {Wang}, \citenamefont {Ying},
  \citenamefont {Zhou}, \citenamefont {Sun}, \citenamefont {Wen}, \citenamefont
  {Zhou}, \citenamefont {Li}, \citenamefont {Zhang}, \citenamefont {Han},
  \citenamefont {Xiao}, \citenamefont {Chow}, \citenamefont {Yang},
  \citenamefont {Struzhkin}, \citenamefont {Zhao},\ and\ \citenamefont
  {Mao}}]{Wang2018}%
  \BibitemOpen
  \bibfield  {author} {\bibinfo {author} {\bibfnamefont {Y.}~\bibnamefont
  {Wang}}, \bibinfo {author} {\bibfnamefont {J.}~\bibnamefont {Ying}}, \bibinfo
  {author} {\bibfnamefont {Z.}~\bibnamefont {Zhou}}, \bibinfo {author}
  {\bibfnamefont {J.}~\bibnamefont {Sun}}, \bibinfo {author} {\bibfnamefont
  {T.}~\bibnamefont {Wen}}, \bibinfo {author} {\bibfnamefont {Y.}~\bibnamefont
  {Zhou}}, \bibinfo {author} {\bibfnamefont {N.}~\bibnamefont {Li}}, \bibinfo
  {author} {\bibfnamefont {Q.}~\bibnamefont {Zhang}}, \bibinfo {author}
  {\bibfnamefont {F.}~\bibnamefont {Han}}, \bibinfo {author} {\bibfnamefont
  {Y.}~\bibnamefont {Xiao}}, \bibinfo {author} {\bibfnamefont {P.}~\bibnamefont
  {Chow}}, \bibinfo {author} {\bibfnamefont {W.}~\bibnamefont {Yang}}, \bibinfo
  {author} {\bibfnamefont {V.~V.}\ \bibnamefont {Struzhkin}}, \bibinfo {author}
  {\bibfnamefont {Y.}~\bibnamefont {Zhao}}, \ and\ \bibinfo {author}
  {\bibfnamefont {H.-k.}\ \bibnamefont {Mao}},\ }\href {\doibase
  10.1038/s41467-018-04326-1} {\bibfield  {journal} {\bibinfo  {journal} {Nat.
  Commun.}\ }\textbf {\bibinfo {volume} {9}},\ \bibinfo {pages} {1914}
  (\bibinfo {year} {2018}{\natexlab{b}})}\BibitemShut {NoStop}%
\bibitem [{\citenamefont {Liu}\ \emph {et~al.}(2021)\citenamefont {Liu},
  \citenamefont {Wang}, \citenamefont {Fu}, \citenamefont {Zhang},
  \citenamefont {Huang}, \citenamefont {Su}, \citenamefont {Lin}, \citenamefont
  {Chen}, \citenamefont {Yu}, \citenamefont {Cui}, \citenamefont {Mei},\ and\
  \citenamefont {Dai}}]{2021_PRB_Liu}%
  \BibitemOpen
  \bibfield  {author} {\bibinfo {author} {\bibfnamefont {Q.}~\bibnamefont
  {Liu}}, \bibinfo {author} {\bibfnamefont {L.}~\bibnamefont {Wang}}, \bibinfo
  {author} {\bibfnamefont {Y.}~\bibnamefont {Fu}}, \bibinfo {author}
  {\bibfnamefont {X.}~\bibnamefont {Zhang}}, \bibinfo {author} {\bibfnamefont
  {L.}~\bibnamefont {Huang}}, \bibinfo {author} {\bibfnamefont
  {H.}~\bibnamefont {Su}}, \bibinfo {author} {\bibfnamefont {J.}~\bibnamefont
  {Lin}}, \bibinfo {author} {\bibfnamefont {X.}~\bibnamefont {Chen}}, \bibinfo
  {author} {\bibfnamefont {D.}~\bibnamefont {Yu}}, \bibinfo {author}
  {\bibfnamefont {X.}~\bibnamefont {Cui}}, \bibinfo {author} {\bibfnamefont
  {J.-W.}\ \bibnamefont {Mei}}, \ and\ \bibinfo {author} {\bibfnamefont
  {J.-F.}\ \bibnamefont {Dai}},\ }\href {\doibase 10.1103/PhysRevB.103.235411}
  {\bibfield  {journal} {\bibinfo  {journal} {Phys. Rev. B}\ }\textbf {\bibinfo
  {volume} {103}},\ \bibinfo {pages} {235411} (\bibinfo {year}
  {2021})}\BibitemShut {NoStop}%
\bibitem [{\citenamefont {Wiedenmann}\ \emph {et~al.}(1981)\citenamefont
  {Wiedenmann}, \citenamefont {Rossat-Mignod}, \citenamefont {Louisy},
  \citenamefont {Brec},\ and\ \citenamefont {Rouxel}}]{12_1981SSC}%
  \BibitemOpen
  \bibfield  {author} {\bibinfo {author} {\bibfnamefont {A.}~\bibnamefont
  {Wiedenmann}}, \bibinfo {author} {\bibfnamefont {J.}~\bibnamefont
  {Rossat-Mignod}}, \bibinfo {author} {\bibfnamefont {A.}~\bibnamefont
  {Louisy}}, \bibinfo {author} {\bibfnamefont {R.}~\bibnamefont {Brec}}, \ and\
  \bibinfo {author} {\bibfnamefont {J.}~\bibnamefont {Rouxel}},\ }\href
  {\doibase https://doi.org/10.1016/0038-1098(81)90253-2} {\bibfield  {journal}
  {\bibinfo  {journal} {Solid State Commun.}\ }\textbf {\bibinfo {volume}
  {40}},\ \bibinfo {pages} {1067} (\bibinfo {year} {1981})}\BibitemShut
  {NoStop}%
\bibitem [{\citenamefont {Grasso}\ and\ \citenamefont
  {Silipigni}(1999)}]{33_XAS}%
  \BibitemOpen
  \bibfield  {author} {\bibinfo {author} {\bibfnamefont {V.}~\bibnamefont
  {Grasso}}\ and\ \bibinfo {author} {\bibfnamefont {L.}~\bibnamefont
  {Silipigni}},\ }\href {\doibase 10.1364/JOSAB.16.000132} {\bibfield
  {journal} {\bibinfo  {journal} {J. Opt. Soc. Am. B}\ }\textbf {\bibinfo
  {volume} {16}},\ \bibinfo {pages} {132} (\bibinfo {year} {1999})}\BibitemShut
  {NoStop}%
\bibitem [{\citenamefont {Li}\ \emph {et~al.}(2014)\citenamefont {Li},
  \citenamefont {Wu},\ and\ \citenamefont {Yang}}]{1_JACS2014}%
  \BibitemOpen
  \bibfield  {author} {\bibinfo {author} {\bibfnamefont {X.}~\bibnamefont
  {Li}}, \bibinfo {author} {\bibfnamefont {X.}~\bibnamefont {Wu}}, \ and\
  \bibinfo {author} {\bibfnamefont {J.}~\bibnamefont {Yang}},\ }\href {\doibase
  10.1021/ja505097m} {\bibfield  {journal} {\bibinfo  {journal} {J. Am. Chem.
  Soc.}\ }\textbf {\bibinfo {volume} {136}},\ \bibinfo {pages} {11065}
  (\bibinfo {year} {2014})}\BibitemShut {NoStop}%
\bibitem [{\citenamefont {Wang}\ \emph {et~al.}(2016)\citenamefont {Wang},
  \citenamefont {Zhou}, \citenamefont {Wen}, \citenamefont {Zhou},
  \citenamefont {Li}, \citenamefont {Han}, \citenamefont {Xiao}, \citenamefont
  {Chow}, \citenamefont {Sun}, \citenamefont {Pravica}, \citenamefont
  {Cornelius}, \citenamefont {Yang},\ and\ \citenamefont
  {Zhao}}]{WangJACS2016}%
  \BibitemOpen
  \bibfield  {author} {\bibinfo {author} {\bibfnamefont {Y.}~\bibnamefont
  {Wang}}, \bibinfo {author} {\bibfnamefont {Z.}~\bibnamefont {Zhou}}, \bibinfo
  {author} {\bibfnamefont {T.}~\bibnamefont {Wen}}, \bibinfo {author}
  {\bibfnamefont {Y.}~\bibnamefont {Zhou}}, \bibinfo {author} {\bibfnamefont
  {N.}~\bibnamefont {Li}}, \bibinfo {author} {\bibfnamefont {F.}~\bibnamefont
  {Han}}, \bibinfo {author} {\bibfnamefont {Y.}~\bibnamefont {Xiao}}, \bibinfo
  {author} {\bibfnamefont {P.}~\bibnamefont {Chow}}, \bibinfo {author}
  {\bibfnamefont {J.}~\bibnamefont {Sun}}, \bibinfo {author} {\bibfnamefont
  {M.}~\bibnamefont {Pravica}}, \bibinfo {author} {\bibfnamefont {A.~L.}\
  \bibnamefont {Cornelius}}, \bibinfo {author} {\bibfnamefont {W.}~\bibnamefont
  {Yang}}, \ and\ \bibinfo {author} {\bibfnamefont {Y.}~\bibnamefont {Zhao}},\
  }\href {\doibase 10.1021/jacs.6b10225} {\bibfield  {journal} {\bibinfo
  {journal} {J. Am. Chem. Soc.}\ }\textbf {\bibinfo {volume} {138}},\ \bibinfo
  {pages} {15751} (\bibinfo {year} {2016})}\BibitemShut {NoStop}%
\bibitem [{\citenamefont {Sivadas}\ \emph {et~al.}(2016)\citenamefont
  {Sivadas}, \citenamefont {Okamoto},\ and\ \citenamefont
  {Xiao}}]{SivadasPRL2016}%
  \BibitemOpen
  \bibfield  {author} {\bibinfo {author} {\bibfnamefont {N.}~\bibnamefont
  {Sivadas}}, \bibinfo {author} {\bibfnamefont {S.}~\bibnamefont {Okamoto}}, \
  and\ \bibinfo {author} {\bibfnamefont {D.}~\bibnamefont {Xiao}},\ }\href
  {\doibase 10.1103/PhysRevLett.117.267203} {\bibfield  {journal} {\bibinfo
  {journal} {Phys. Rev. Lett.}\ }\textbf {\bibinfo {volume} {117}},\ \bibinfo
  {pages} {267203} (\bibinfo {year} {2016})}\BibitemShut {NoStop}%
\bibitem [{\citenamefont {Onga}\ \emph {et~al.}(2020)\citenamefont {Onga},
  \citenamefont {Sugita}, \citenamefont {Ideue}, \citenamefont {Nakagawa},
  \citenamefont {Suzuki}, \citenamefont {Motome},\ and\ \citenamefont
  {Iwasa}}]{61_Onga2020}%
  \BibitemOpen
  \bibfield  {author} {\bibinfo {author} {\bibfnamefont {M.}~\bibnamefont
  {Onga}}, \bibinfo {author} {\bibfnamefont {Y.}~\bibnamefont {Sugita}},
  \bibinfo {author} {\bibfnamefont {T.}~\bibnamefont {Ideue}}, \bibinfo
  {author} {\bibfnamefont {Y.}~\bibnamefont {Nakagawa}}, \bibinfo {author}
  {\bibfnamefont {R.}~\bibnamefont {Suzuki}}, \bibinfo {author} {\bibfnamefont
  {Y.}~\bibnamefont {Motome}}, \ and\ \bibinfo {author} {\bibfnamefont
  {Y.}~\bibnamefont {Iwasa}},\ }\href {\doibase 10.1021/acs.nanolett.0c01493}
  {\bibfield  {journal} {\bibinfo  {journal} {Nano Lett.}\ }\textbf {\bibinfo
  {volume} {20}},\ \bibinfo {pages} {4625} (\bibinfo {year}
  {2020})}\BibitemShut {NoStop}%
\bibitem [{\citenamefont {Ni}\ \emph {et~al.}(2021)\citenamefont {Ni},
  \citenamefont {Haglund}, \citenamefont {Wang}, \citenamefont {Xu},
  \citenamefont {Bernhard}, \citenamefont {Mandrus}, \citenamefont {Qian},
  \citenamefont {Mele}, \citenamefont {Kane},\ and\ \citenamefont
  {Wu}}]{11_naturenano2021}%
  \BibitemOpen
  \bibfield  {author} {\bibinfo {author} {\bibfnamefont {Z.}~\bibnamefont
  {Ni}}, \bibinfo {author} {\bibfnamefont {A.~V.}\ \bibnamefont {Haglund}},
  \bibinfo {author} {\bibfnamefont {H.}~\bibnamefont {Wang}}, \bibinfo {author}
  {\bibfnamefont {B.}~\bibnamefont {Xu}}, \bibinfo {author} {\bibfnamefont
  {C.}~\bibnamefont {Bernhard}}, \bibinfo {author} {\bibfnamefont {D.~G.}\
  \bibnamefont {Mandrus}}, \bibinfo {author} {\bibfnamefont {X.}~\bibnamefont
  {Qian}}, \bibinfo {author} {\bibfnamefont {E.~J.}\ \bibnamefont {Mele}},
  \bibinfo {author} {\bibfnamefont {C.~L.}\ \bibnamefont {Kane}}, \ and\
  \bibinfo {author} {\bibfnamefont {L.}~\bibnamefont {Wu}},\ }\href {\doibase
  10.1038/s41565-021-00885-5} {\bibfield  {journal} {\bibinfo  {journal} {Nat.
  Nanotechnol.}\ }\textbf {\bibinfo {volume} {16}},\ \bibinfo {pages} {782}
  (\bibinfo {year} {2021})}\BibitemShut {NoStop}%
\bibitem [{\citenamefont {Calder}\ \emph {et~al.}(2021)\citenamefont {Calder},
  \citenamefont {Haglund}, \citenamefont {Kolesnikov},\ and\ \citenamefont
  {Mandrus}}]{ref26_PRB}%
  \BibitemOpen
  \bibfield  {author} {\bibinfo {author} {\bibfnamefont {S.}~\bibnamefont
  {Calder}}, \bibinfo {author} {\bibfnamefont {A.~V.}\ \bibnamefont {Haglund}},
  \bibinfo {author} {\bibfnamefont {A.~I.}\ \bibnamefont {Kolesnikov}}, \ and\
  \bibinfo {author} {\bibfnamefont {D.}~\bibnamefont {Mandrus}},\ }\href
  {\doibase 10.1103/PhysRevB.103.024414} {\bibfield  {journal} {\bibinfo
  {journal} {Phys. Rev. B}\ }\textbf {\bibinfo {volume} {103}},\ \bibinfo
  {pages} {024414} (\bibinfo {year} {2021})}\BibitemShut {NoStop}%
\bibitem [{\citenamefont {Bhutani}\ \emph {et~al.}(2020)\citenamefont
  {Bhutani}, \citenamefont {Zuo}, \citenamefont {McAuliffe}, \citenamefont
  {dela Cruz},\ and\ \citenamefont {Shoemaker}}]{ref37_PRM}%
  \BibitemOpen
  \bibfield  {author} {\bibinfo {author} {\bibfnamefont {A.}~\bibnamefont
  {Bhutani}}, \bibinfo {author} {\bibfnamefont {J.~L.}\ \bibnamefont {Zuo}},
  \bibinfo {author} {\bibfnamefont {R.~D.}\ \bibnamefont {McAuliffe}}, \bibinfo
  {author} {\bibfnamefont {C.~R.}\ \bibnamefont {dela Cruz}}, \ and\ \bibinfo
  {author} {\bibfnamefont {D.~P.}\ \bibnamefont {Shoemaker}},\ }\href {\doibase
  10.1103/PhysRevMaterials.4.034411} {\bibfield  {journal} {\bibinfo  {journal}
  {Phys. Rev. Mater.}\ }\textbf {\bibinfo {volume} {4}},\ \bibinfo {pages}
  {034411} (\bibinfo {year} {2020})}\BibitemShut {NoStop}%
\bibitem [{\citenamefont {Mermin}\ and\ \citenamefont {Wagner}(1966)}]{Mermin}%
  \BibitemOpen
  \bibfield  {author} {\bibinfo {author} {\bibfnamefont {N.~D.}\ \bibnamefont
  {Mermin}}\ and\ \bibinfo {author} {\bibfnamefont {H.}~\bibnamefont
  {Wagner}},\ }\href {\doibase 10.1103/PhysRevLett.17.1133} {\bibfield
  {journal} {\bibinfo  {journal} {Phys. Rev. Lett.}\ }\textbf {\bibinfo
  {volume} {17}},\ \bibinfo {pages} {1133} (\bibinfo {year}
  {1966})}\BibitemShut {NoStop}%
\bibitem [{\citenamefont {Kresse}\ and\ \citenamefont
  {Furthm\"uller}(1996)}]{VASP}%
  \BibitemOpen
  \bibfield  {author} {\bibinfo {author} {\bibfnamefont {G.}~\bibnamefont
  {Kresse}}\ and\ \bibinfo {author} {\bibfnamefont {J.}~\bibnamefont
  {Furthm\"uller}},\ }\href {\doibase 10.1103/PhysRevB.54.11169} {\bibfield
  {journal} {\bibinfo  {journal} {Phys. Rev. B}\ }\textbf {\bibinfo {volume}
  {54}},\ \bibinfo {pages} {11169} (\bibinfo {year} {1996})}\BibitemShut
  {NoStop}%
\bibitem [{\citenamefont {Kresse}\ and\ \citenamefont {Joubert}(1999)}]{PAW}%
  \BibitemOpen
  \bibfield  {author} {\bibinfo {author} {\bibfnamefont {G.}~\bibnamefont
  {Kresse}}\ and\ \bibinfo {author} {\bibfnamefont {D.}~\bibnamefont
  {Joubert}},\ }\href {\doibase 10.1103/PhysRevB.59.1758} {\bibfield  {journal}
  {\bibinfo  {journal} {Phys. Rev. B}\ }\textbf {\bibinfo {volume} {59}},\
  \bibinfo {pages} {1758} (\bibinfo {year} {1999})}\BibitemShut {NoStop}%
\bibitem [{\citenamefont {Perdew}\ \emph {et~al.}(1996)\citenamefont {Perdew},
  \citenamefont {Burke},\ and\ \citenamefont {Ernzerhof}}]{PBE}%
  \BibitemOpen
  \bibfield  {author} {\bibinfo {author} {\bibfnamefont {J.~P.}\ \bibnamefont
  {Perdew}}, \bibinfo {author} {\bibfnamefont {K.}~\bibnamefont {Burke}}, \
  and\ \bibinfo {author} {\bibfnamefont {M.}~\bibnamefont {Ernzerhof}},\ }\href
  {\doibase 10.1103/PhysRevLett.77.3865} {\bibfield  {journal} {\bibinfo
  {journal} {Phys. Rev. Lett.}\ }\textbf {\bibinfo {volume} {77}},\ \bibinfo
  {pages} {3865} (\bibinfo {year} {1996})}\BibitemShut {NoStop}%
\bibitem [{\citenamefont {Monkhorst}\ and\ \citenamefont
  {Pack}(1976)}]{MP_Method}%
  \BibitemOpen
  \bibfield  {author} {\bibinfo {author} {\bibfnamefont {H.~J.}\ \bibnamefont
  {Monkhorst}}\ and\ \bibinfo {author} {\bibfnamefont {J.~D.}\ \bibnamefont
  {Pack}},\ }\href {\doibase 10.1103/PhysRevB.13.5188} {\bibfield  {journal}
  {\bibinfo  {journal} {Phys. Rev. B}\ }\textbf {\bibinfo {volume} {13}},\
  \bibinfo {pages} {5188} (\bibinfo {year} {1976})}\BibitemShut {NoStop}%
\bibitem [{\citenamefont {Dudarev}\ \emph {et~al.}(1998)\citenamefont
  {Dudarev}, \citenamefont {Botton}, \citenamefont {Savrasov}, \citenamefont
  {Humphreys},\ and\ \citenamefont {Sutton}}]{UTYPE2}%
  \BibitemOpen
  \bibfield  {author} {\bibinfo {author} {\bibfnamefont {S.~L.}\ \bibnamefont
  {Dudarev}}, \bibinfo {author} {\bibfnamefont {G.~A.}\ \bibnamefont {Botton}},
  \bibinfo {author} {\bibfnamefont {S.~Y.}\ \bibnamefont {Savrasov}}, \bibinfo
  {author} {\bibfnamefont {C.~J.}\ \bibnamefont {Humphreys}}, \ and\ \bibinfo
  {author} {\bibfnamefont {A.~P.}\ \bibnamefont {Sutton}},\ }\href {\doibase
  10.1103/PhysRevB.57.1505} {\bibfield  {journal} {\bibinfo  {journal} {Phys.
  Rev. B}\ }\textbf {\bibinfo {volume} {57}},\ \bibinfo {pages} {1505}
  (\bibinfo {year} {1998})}\BibitemShut {NoStop}%
\bibitem [{\citenamefont {Li}\ \emph {et~al.}(2019{\natexlab{b}})\citenamefont
  {Li}, \citenamefont {Li}, \citenamefont {Du}, \citenamefont {Wang},
  \citenamefont {Gu}, \citenamefont {Zhang}, \citenamefont {He}, \citenamefont
  {Duan},\ and\ \citenamefont {Xu}}]{U_SciAdv_2019}%
  \BibitemOpen
  \bibfield  {author} {\bibinfo {author} {\bibfnamefont {J.}~\bibnamefont
  {Li}}, \bibinfo {author} {\bibfnamefont {Y.}~\bibnamefont {Li}}, \bibinfo
  {author} {\bibfnamefont {S.}~\bibnamefont {Du}}, \bibinfo {author}
  {\bibfnamefont {Z.}~\bibnamefont {Wang}}, \bibinfo {author} {\bibfnamefont
  {B.-L.}\ \bibnamefont {Gu}}, \bibinfo {author} {\bibfnamefont {S.-C.}\
  \bibnamefont {Zhang}}, \bibinfo {author} {\bibfnamefont {K.}~\bibnamefont
  {He}}, \bibinfo {author} {\bibfnamefont {W.}~\bibnamefont {Duan}}, \ and\
  \bibinfo {author} {\bibfnamefont {Y.}~\bibnamefont {Xu}},\ }\href {\doibase
  10.1126/sciadv.aaw5685} {\bibfield  {journal} {\bibinfo  {journal} {Sci.
  Adv.}\ }\textbf {\bibinfo {volume} {5}},\ \bibinfo {pages} {eaaw5685}
  (\bibinfo {year} {2019}{\natexlab{b}})}\BibitemShut {NoStop}%
\bibitem [{\citenamefont {An}\ \emph {et~al.}(2021)\citenamefont {An},
  \citenamefont {Wang}, \citenamefont {Gong}, \citenamefont {Hou},
  \citenamefont {Ma}, \citenamefont {Zhu}, \citenamefont {Zhao}, \citenamefont
  {Wang}, \citenamefont {Ma}, \citenamefont {Wang}, \citenamefont {Wu},\ and\
  \citenamefont {Liu}}]{U_npj_2021}%
  \BibitemOpen
  \bibfield  {author} {\bibinfo {author} {\bibfnamefont {Y.}~\bibnamefont
  {An}}, \bibinfo {author} {\bibfnamefont {K.}~\bibnamefont {Wang}}, \bibinfo
  {author} {\bibfnamefont {S.}~\bibnamefont {Gong}}, \bibinfo {author}
  {\bibfnamefont {Y.}~\bibnamefont {Hou}}, \bibinfo {author} {\bibfnamefont
  {C.}~\bibnamefont {Ma}}, \bibinfo {author} {\bibfnamefont {M.}~\bibnamefont
  {Zhu}}, \bibinfo {author} {\bibfnamefont {C.}~\bibnamefont {Zhao}}, \bibinfo
  {author} {\bibfnamefont {T.}~\bibnamefont {Wang}}, \bibinfo {author}
  {\bibfnamefont {S.}~\bibnamefont {Ma}}, \bibinfo {author} {\bibfnamefont
  {H.}~\bibnamefont {Wang}}, \bibinfo {author} {\bibfnamefont {R.}~\bibnamefont
  {Wu}}, \ and\ \bibinfo {author} {\bibfnamefont {W.}~\bibnamefont {Liu}},\
  }\href {\doibase 10.1038/s41524-021-00513-9} {\bibfield  {journal} {\bibinfo
  {journal} {NPJ Comput. Mater.}\ }\textbf {\bibinfo {volume} {7}},\ \bibinfo
  {pages} {45} (\bibinfo {year} {2021})}\BibitemShut {NoStop}%
\bibitem [{\citenamefont {Wu}\ \emph {et~al.}(2009)\citenamefont {Wu},
  \citenamefont {Burnus}, \citenamefont {Hu}, \citenamefont {Martin},
  \citenamefont {Maignan}, \citenamefont {Cezar}, \citenamefont {Tanaka},
  \citenamefont {Brookes}, \citenamefont {Khomskii},\ and\ \citenamefont
  {Tjeng}}]{U_PRL_2009}%
  \BibitemOpen
  \bibfield  {author} {\bibinfo {author} {\bibfnamefont {H.}~\bibnamefont
  {Wu}}, \bibinfo {author} {\bibfnamefont {T.}~\bibnamefont {Burnus}}, \bibinfo
  {author} {\bibfnamefont {Z.}~\bibnamefont {Hu}}, \bibinfo {author}
  {\bibfnamefont {C.}~\bibnamefont {Martin}}, \bibinfo {author} {\bibfnamefont
  {A.}~\bibnamefont {Maignan}}, \bibinfo {author} {\bibfnamefont {J.~C.}\
  \bibnamefont {Cezar}}, \bibinfo {author} {\bibfnamefont {A.}~\bibnamefont
  {Tanaka}}, \bibinfo {author} {\bibfnamefont {N.~B.}\ \bibnamefont {Brookes}},
  \bibinfo {author} {\bibfnamefont {D.~I.}\ \bibnamefont {Khomskii}}, \ and\
  \bibinfo {author} {\bibfnamefont {L.~H.}\ \bibnamefont {Tjeng}},\ }\href
  {\doibase 10.1103/PhysRevLett.102.026404} {\bibfield  {journal} {\bibinfo
  {journal} {Phys. Rev. Lett.}\ }\textbf {\bibinfo {volume} {102}},\ \bibinfo
  {pages} {026404} (\bibinfo {year} {2009})}\BibitemShut {NoStop}%
\bibitem [{\citenamefont {Mostofi}\ \emph {et~al.}(2008)\citenamefont
  {Mostofi}, \citenamefont {Yates}, \citenamefont {Lee}, \citenamefont {Souza},
  \citenamefont {Vanderbilt},\ and\ \citenamefont {Marzari}}]{Wannier90}%
  \BibitemOpen
  \bibfield  {author} {\bibinfo {author} {\bibfnamefont {A.~A.}\ \bibnamefont
  {Mostofi}}, \bibinfo {author} {\bibfnamefont {J.~R.}\ \bibnamefont {Yates}},
  \bibinfo {author} {\bibfnamefont {Y.-S.}\ \bibnamefont {Lee}}, \bibinfo
  {author} {\bibfnamefont {I.}~\bibnamefont {Souza}}, \bibinfo {author}
  {\bibfnamefont {D.}~\bibnamefont {Vanderbilt}}, \ and\ \bibinfo {author}
  {\bibfnamefont {N.}~\bibnamefont {Marzari}},\ }\href {\doibase
  https://doi.org/10.1016/j.cpc.2007.11.016} {\bibfield  {journal} {\bibinfo
  {journal} {Comput. Phys. Commun.}\ }\textbf {\bibinfo {volume} {178}},\
  \bibinfo {pages} {685} (\bibinfo {year} {2008})}\BibitemShut {NoStop}%
\bibitem [{\citenamefont {Marzari}\ \emph {et~al.}(2012)\citenamefont
  {Marzari}, \citenamefont {Mostofi}, \citenamefont {Yates}, \citenamefont
  {Souza},\ and\ \citenamefont {Vanderbilt}}]{RMP_MLWF}%
  \BibitemOpen
  \bibfield  {author} {\bibinfo {author} {\bibfnamefont {N.}~\bibnamefont
  {Marzari}}, \bibinfo {author} {\bibfnamefont {A.~A.}\ \bibnamefont
  {Mostofi}}, \bibinfo {author} {\bibfnamefont {J.~R.}\ \bibnamefont {Yates}},
  \bibinfo {author} {\bibfnamefont {I.}~\bibnamefont {Souza}}, \ and\ \bibinfo
  {author} {\bibfnamefont {D.}~\bibnamefont {Vanderbilt}},\ }\href {\doibase
  10.1103/RevModPhys.84.1419} {\bibfield  {journal} {\bibinfo  {journal} {Rev.
  Mod. Phys.}\ }\textbf {\bibinfo {volume} {84}},\ \bibinfo {pages} {1419}
  (\bibinfo {year} {2012})}\BibitemShut {NoStop}%
\bibitem [{\citenamefont {Metropolis}\ and\ \citenamefont
  {S.Ulam}(1949)}]{MC_Metropolis}%
  \BibitemOpen
  \bibfield  {author} {\bibinfo {author} {\bibfnamefont {N.}~\bibnamefont
  {Metropolis}}\ and\ \bibinfo {author} {\bibnamefont {S.Ulam}},\ }\href
  {\doibase 10.1080/01621459.1949.10483310} {\bibfield  {journal} {\bibinfo
  {journal} {J. Am. Stat. Assoc.}\ }\textbf {\bibinfo {volume} {44}},\ \bibinfo
  {pages} {335} (\bibinfo {year} {1949})}\BibitemShut {NoStop}%
\bibitem [{\citenamefont {Fujii}\ \emph {et~al.}(2022)\citenamefont {Fujii},
  \citenamefont {Yamaguchi}, \citenamefont {Ohkochi}, \citenamefont {De},
  \citenamefont {Cheong},\ and\ \citenamefont
  {Mizokawa}}]{ref64_PRB2022_masato}%
  \BibitemOpen
  \bibfield  {author} {\bibinfo {author} {\bibfnamefont {M.}~\bibnamefont
  {Fujii}}, \bibinfo {author} {\bibfnamefont {T.}~\bibnamefont {Yamaguchi}},
  \bibinfo {author} {\bibfnamefont {T.}~\bibnamefont {Ohkochi}}, \bibinfo
  {author} {\bibfnamefont {C.}~\bibnamefont {De}}, \bibinfo {author}
  {\bibfnamefont {S.-W.}\ \bibnamefont {Cheong}}, \ and\ \bibinfo {author}
  {\bibfnamefont {T.}~\bibnamefont {Mizokawa}},\ }\href {\doibase
  10.1103/PhysRevB.106.035118} {\bibfield  {journal} {\bibinfo  {journal}
  {Phys. Rev. B}\ }\textbf {\bibinfo {volume} {106}},\ \bibinfo {pages}
  {035118} (\bibinfo {year} {2022})}\BibitemShut {NoStop}%
\bibitem [{SM()}]{SM}%
  \BibitemOpen
  \href@noop {} {\bibinfo  {journal} {See Supplemental Materials at XXX... for
  the calculation of the magnetic exchange parameters, the hopping integrals of
  spin-up channels, the distributions of the MAE in the FM state over the
  reciprocal space, and the polar diagrams of the MAE under strains}\
  }\BibitemShut {NoStop}%
\bibitem [{\citenamefont {Khomskii}(2014)}]{Khomskii_2014}%
  \BibitemOpen
\bibfield  {journal} {  }\bibfield  {author} {\bibinfo {author} {\bibfnamefont
  {D.~I.}\ \bibnamefont {Khomskii}},\ }\href {\doibase
  10.1017/CBO9781139096782} {\emph {\bibinfo {title} {Transition Metal
  Compounds}}}\ (\bibinfo  {publisher} {Cambridge University Press},\ \bibinfo
  {year} {2014})\BibitemShut {NoStop}%
\bibitem [{\citenamefont {Yang}\ \emph {et~al.}(2020)\citenamefont {Yang},
  \citenamefont {Fan}, \citenamefont {Wang}, \citenamefont {Khomskii},\ and\
  \citenamefont {Wu}}]{VI3}%
  \BibitemOpen
  \bibfield  {author} {\bibinfo {author} {\bibfnamefont {K.}~\bibnamefont
  {Yang}}, \bibinfo {author} {\bibfnamefont {F.}~\bibnamefont {Fan}}, \bibinfo
  {author} {\bibfnamefont {H.}~\bibnamefont {Wang}}, \bibinfo {author}
  {\bibfnamefont {D.~I.}\ \bibnamefont {Khomskii}}, \ and\ \bibinfo {author}
  {\bibfnamefont {H.}~\bibnamefont {Wu}},\ }\href {\doibase
  10.1103/PhysRevB.101.100402} {\bibfield  {journal} {\bibinfo  {journal}
  {Phys. Rev. B}\ }\textbf {\bibinfo {volume} {101}},\ \bibinfo {pages}
  {100402(R)} (\bibinfo {year} {2020})}\BibitemShut {NoStop}%
\bibitem [{\citenamefont {Yang}\ \emph {et~al.}(2021)\citenamefont {Yang},
  \citenamefont {Wang}, \citenamefont {Liu}, \citenamefont {Lu},\ and\
  \citenamefont {Wu}}]{CrSBr}%
  \BibitemOpen
  \bibfield  {author} {\bibinfo {author} {\bibfnamefont {K.}~\bibnamefont
  {Yang}}, \bibinfo {author} {\bibfnamefont {G.}~\bibnamefont {Wang}}, \bibinfo
  {author} {\bibfnamefont {L.}~\bibnamefont {Liu}}, \bibinfo {author}
  {\bibfnamefont {D.}~\bibnamefont {Lu}}, \ and\ \bibinfo {author}
  {\bibfnamefont {H.}~\bibnamefont {Wu}},\ }\href {\doibase
  10.1103/PhysRevB.104.144416} {\bibfield  {journal} {\bibinfo  {journal}
  {Phys. Rev. B}\ }\textbf {\bibinfo {volume} {104}},\ \bibinfo {pages}
  {144416} (\bibinfo {year} {2021})}\BibitemShut {NoStop}%
\bibitem [{\citenamefont {Wang}\ \emph {et~al.}(2021)\citenamefont {Wang},
  \citenamefont {Liu}, \citenamefont {Yang},\ and\ \citenamefont
  {Wu}}]{CrSbSe}%
  \BibitemOpen
  \bibfield  {author} {\bibinfo {author} {\bibfnamefont {G.}~\bibnamefont
  {Wang}}, \bibinfo {author} {\bibfnamefont {L.}~\bibnamefont {Liu}}, \bibinfo
  {author} {\bibfnamefont {K.}~\bibnamefont {Yang}}, \ and\ \bibinfo {author}
  {\bibfnamefont {H.}~\bibnamefont {Wu}},\ }\href {\doibase
  10.1103/PhysRevMaterials.5.124412} {\bibfield  {journal} {\bibinfo  {journal}
  {Phys. Rev. Mater.}\ }\textbf {\bibinfo {volume} {5}},\ \bibinfo {pages}
  {124412} (\bibinfo {year} {2021})}\BibitemShut {NoStop}%
\bibitem [{\citenamefont {Wang}\ \emph {et~al.}(1993)\citenamefont {Wang},
  \citenamefont {Wu},\ and\ \citenamefont {Freeman}}]{PhysRevLett.70.869}%
  \BibitemOpen
  \bibfield  {author} {\bibinfo {author} {\bibfnamefont {D.-s.}\ \bibnamefont
  {Wang}}, \bibinfo {author} {\bibfnamefont {R.}~\bibnamefont {Wu}}, \ and\
  \bibinfo {author} {\bibfnamefont {A.~J.}\ \bibnamefont {Freeman}},\ }\href
  {\doibase 10.1103/PhysRevLett.70.869} {\bibfield  {journal} {\bibinfo
  {journal} {Phys. Rev. Lett.}\ }\textbf {\bibinfo {volume} {70}},\ \bibinfo
  {pages} {869} (\bibinfo {year} {1993})}\BibitemShut {NoStop}%
\bibitem [{\citenamefont {van~der Laan}(1998)}]{Laan_1998_JPCM}%
  \BibitemOpen
  \bibfield  {author} {\bibinfo {author} {\bibfnamefont {G.}~\bibnamefont
  {van~der Laan}},\ }\href {\doibase 10.1088/0953-8984/10/14/012} {\bibfield
  {journal} {\bibinfo  {journal} {J. Phys.: Condens. Matter}\ }\textbf
  {\bibinfo {volume} {10}},\ \bibinfo {pages} {3239} (\bibinfo {year}
  {1998})}\BibitemShut {NoStop}%
\bibitem [{\citenamefont {Qin}\ \emph {et~al.}(2017)\citenamefont {Qin},
  \citenamefont {Qin}, \citenamefont {Shao},\ and\ \citenamefont
  {Zuo}}]{Qin_Nanoscale}%
  \BibitemOpen
  \bibfield  {author} {\bibinfo {author} {\bibfnamefont {Z.}~\bibnamefont
  {Qin}}, \bibinfo {author} {\bibfnamefont {G.}~\bibnamefont {Qin}}, \bibinfo
  {author} {\bibfnamefont {B.}~\bibnamefont {Shao}}, \ and\ \bibinfo {author}
  {\bibfnamefont {X.}~\bibnamefont {Zuo}},\ }\href {\doibase
  10.1039/C7NR03164E} {\bibfield  {journal} {\bibinfo  {journal} {Nanoscale}\
  }\textbf {\bibinfo {volume} {9}},\ \bibinfo {pages} {11657} (\bibinfo {year}
  {2017})}\BibitemShut {NoStop}%
\bibitem [{\citenamefont {Yue}\ \emph {et~al.}(2020)\citenamefont {Yue},
  \citenamefont {Jiang}, \citenamefont {Han}, \citenamefont {Wang},
  \citenamefont {Ren},\ and\ \citenamefont {Wu}}]{YueJMMM}%
  \BibitemOpen
  \bibfield  {author} {\bibinfo {author} {\bibfnamefont {Y.}~\bibnamefont
  {Yue}}, \bibinfo {author} {\bibfnamefont {C.}~\bibnamefont {Jiang}}, \bibinfo
  {author} {\bibfnamefont {Y.}~\bibnamefont {Han}}, \bibinfo {author}
  {\bibfnamefont {M.}~\bibnamefont {Wang}}, \bibinfo {author} {\bibfnamefont
  {J.}~\bibnamefont {Ren}}, \ and\ \bibinfo {author} {\bibfnamefont
  {Y.}~\bibnamefont {Wu}},\ }\href {\doibase
  https://doi.org/10.1016/j.jmmm.2019.165929} {\bibfield  {journal} {\bibinfo
  {journal} {J. Magn. Magn. Mater.}\ }\textbf {\bibinfo {volume} {496}},\
  \bibinfo {pages} {165929} (\bibinfo {year} {2020})}\BibitemShut {NoStop}%
\end{thebibliography}%
\end{document}